\documentclass[onecolumn]{aa}

\usepackage{graphicx}   
\usepackage{dcolumn}    
\usepackage{bm}         
\usepackage{verbatim}   
\usepackage{booktabs}
\usepackage{amsmath,amssymb}
\usepackage{amsfonts}
\usepackage{color}
\usepackage{natbib}
\usepackage{textcomp}
\usepackage{etoolbox}
\usepackage{multirow}
\usepackage{inputenc}
\usepackage{txfonts}
\bibpunct{(}{)}{;}{a}{}{,}
\usepackage[caption=false]{subfig}
\usepackage{savesym}
\savesymbol{tablenum}
\usepackage{siunitx}
\restoresymbol{SIX}{tablenum}
\usepackage{enumitem}   
\usepackage{xcolor}
\usepackage[normalem]{ulem}
\newcommand\bsout{\bgroup\markoverwith{\textcolor{blue}{\rule[0.5ex]{2pt}{0.4pt}}}\ULon}

\begin{document}
\title{Particle heating and acceleration by reconnecting and non-reconnecting Current Sheets}
\titlerunning{Reconnecting and Non-Reconnecting Current Sheets}
\author{Nikos Sioulas \inst{1,2}, Heinz Isliker \inst{1}, and Loukas Vlahos \inst{1}}
\authorrunning{N. Sioulas et al.}
\institute{$^{1}$ Department of Physics, \; Aristotle University of Thessaloniki\\
GR-52124 Thessaloniki, Greece \\
$^{2}$ Earth, Planetary, and Space Sciences, \;
University of California, Los Angeles \\
Los Angeles, CA 90095, USA}

\date{}
\abstract
{In this article, we study the physics of charged particle energization inside a strongly turbulent plasma, where current sheets naturally appear in evolving large-scale magnetic topologies, but they are split into two populations of \textbf{fractally} distributed reconnecting and non-reconnecting current sheets (CS). In particular, we implement a Monte Carlo simulation to analyze the effects of the fractality and we study how the synergy of
energization at reconnecting CSs and at non-reconnecting CSs
affects the heating, the power-law high energy tail, the escape time, and the acceleration time of electrons and ions. The reconnecting current sheets (RCS) systematically accelerate particles and play a key role in the formation of the power-law tail in energy distributions. On the other hand, the stochastic energization of particles through their interaction with non-reconnecting CSs can account for the heating of the solar corona and the impulsive heating during solar flares.  The combination of the two acceleration mechanisms (stochastic and systematic), commonly present in many explosive events of various sizes, influences the steady-state energy distribution, as well as the transport properties of the particles in position- and energy-space. Our results also suggest that the heating and acceleration characteristics of ions and electrons are similar, the only difference being the time scales required to reach a steady state.}
\maketitle
\section{Introduction}\label{intro}
Magnetic reconnection is a topological reconfiguration of the magnetic field in a plasma, accompanied by plasma heating and particle acceleration \citep{Parker57, Petschek64, BiskampBook2000}. Many astrophysical and laboratory phenomena are closely related to magnetic reconnection e.g.\ solar flares, coronal mass ejection, geomagnetic storms,  sawtooth crash, and Edge Localized Modes (ELMs) in tokamaks, etc. The two-dimensional (2D) plasmoid theory dominates modern reconnection studies \citep{Zwebel16, Loureiro16}. Almost all  2D analytical or numerical studies start with a cartoon with well-formed current sheets (CSs) and follow their evolution. The initiation of a multi-island environment is a key element in the evolution of the 2D reconnecting CS \citep{Coppi6,Shibata01,Drake06}. From 2D  numerical simulations,  particle acceleration by multi-island magnetic reconnection was considered plausible for the energization of particles in many explosive phenomena in space and laboratory plasmas. 2D MHD and Particle in Cell (PIC) simulations analyzing reconnection for mildly relativistic plasma are unable to produce a power-law shaped energy distribution of particles \citep{Che19,Dahlin20}. 

In astrophysical and laboratory plasmas, two topics should be addressed before we start discussing the problem of particle acceleration and heating: (1) the evolution of three dimensional (3D) magnetic reconnection and (2) the fact that the formation of a CS or multiple CSs cannot be decoupled from the subsequent reconnection process, as it is commonly assumed. 

Magnetic reconnection is inherently a 3D process \citep{Boozer19}. Three-dimensional magnetic reconnection from a large-scale CS in the presence of injected and subsequently self-generated turbulence during its evolution is an active field of research with the use of MHD and PIC simulations.  The main result from the analysis of 3D magnetic reconnection is the fragmentation of the initial CS and the formation of multiple reconnection sites \citep{Parnell11}. The reconnection generated current filaments span and interact with each other across the whole initial CS, which results in global magnetic energy consumption. The characteristic of the inflow turbulence driving the large scale CS  forces the field lines to wander around inside the whole initial structure and to gradually release energy by continuously developing new reconnection sites at different positions \citep{Matthaeus86,Lazarian99, Onofri04,Kowal09,Oishi15,Kowal17,Dahlin17,Wang19, Leake20, Rueda21}.  The turbulence-driven initial large-scale CS fragments and expands, filling a large-scale structure with multiple CSs of different characteristic sizes \citep{Isliker19}. We will call this environment in the rest of the article {\bf{turbulent reconnection volume}}  (see more details in the review  \cite{Lazarian20}).

The formation of a large-scale 3D current sheet inside a large-scale turbulent environment can be present in many locations in the heliosphere, e.g.\ the emerging magnetic flux in the solar atmosphere  \citep{Archontis19}, the non-linear evolution of a 3D eruptive magnetic topology leading to Coronal Mass Ejection (CME)  \citep{Inoue18,Jiang16}, or the 3D evolution of the Magneto-tail \citep{Kozak18,Sitnov19, Lu20}, and the CS formed in the Earths magnetopause \citep{Cassak16}. In all the environments mentioned, the large-scale CS is formed and driven by turbulent flows, which gradually lead to a relatively large-scale turbulent reconnection volume \citep{Cheng18,  Chitta20}. 

Another avenue to reach a 3D turbulent reconnection volume is studying the formation of CSs inside 3D MHD turbulence \citep{Dmitruk04,Zhdankin13}. Examples for the spontaneous formation of  CSs inside turbulence are reported in the solar wind \citep{Osman14} and the random shuffling of the emerged magnetic fields in the solar corona by the solar convection zone \citep{Galsgaard96, Galsgaard97a, Gasgaard97b,Rappazzo10, Dahlburg16, Rappazzo13, Kanella17, Kanella18, Einaudi21}.

We can then conclude that, either starting by forming a large scale 3D CS inside an MHD turbulent plasma, or starting with 3D MHD turbulence with spontaneous formation of CSs, a turbulent reconnection environment will be the final state (see the reviews on this topic \cite{Matthaeus11, Cargill12,Lazarian12,Karimabadi13a, Karimabadi03b, Karibabadi2013c,Karimabadi2014}). Inside the turbulent reconnection volume, not all CSs formed will reconnect. As a result, the turbulent reconnection volume will consist of a mixture of reconnecting and non reconnecting CSs \citep{Phan20}.

Large scale magnetic disturbances and coherent structures in fully developed turbulence follow monofractal or multifractal scalings, both in space and laboratory plasmas \citep{Tu95,Shivamoggi97,Biskamp03,Dimitropoulou13,Leonardis13,Schaffner15,Isliker19}. In particular, the CSs inside a turbulent reconnection volume are fractally distributed in space \citep{Isliker19}.

Particle acceleration in 3D turbulent reconnection has been analyzed using several approaches and methods (\cite{Arzner04,Onofri06, Arzner06, Turkmani05, Li19} and see the reviews \cite{Vlahos08Tut, Cargill12, Vlahos19}).  All the reported articles  concluded that 3D magnetic reconnection is responsible for the formation of a power-law, high energy tail of electrons and ions during explosive events.  The power-law index and the maximum energy depend on the method used to analyze the acceleration process and the characteristics of the acceleration volume.

Observations of heating and acceleration of electrons in solar flares suggest that during the explosive phase the energized particles 
are also heated impulsively inside the acceleration volume \citep{Lin03}. \cite{ChenB21} combine microwave and X-ray spectroscopic data from a solar explosion above a bright flare arcade to conclude that the best fit electron energy distribution consists of a thermal ``core" with temperature 25 MK and a non-thermal tail joining the thermal core at 16 KeV with a spectral index of 3.6.
Currently, very little theoretical work has been done on the impulsive (collisionless) heating of the plasma during solar
flares. Current observations report intense preflare heating without particle acceleration, variations in the synergy of impulsive heating and particle acceleration, and particle acceleration without impulsive heating \citep{ChittaN20,Hudson21}. In this article, we propose a mechanism that can accommodate these variations in impulsive heating and particle acceleration in solar flares.

Ion heating and acceleration has been analyzed up to now with simulations utilizing kinetic Alfven wave turbulence associated with isolated reconnecting current sheets, or the pick-up behavior during the entry into reconnection exhausts \citep{ Knizhnik11,Drake14, Liang17, Kumar17, Kunz19}. In other words, the electrons and ions need different energization processes, for which it is not very clear how they are connected with isolated reconnecting current sheets. We show in this article that electrons and ions are accelerated simultaneously in the same turbulent reconnection environment and by the same mechanism.

The formation of CSs in the quiet Sun's magnetic network through the tangling of coronal magnetic field lines by photospheric flows was proposed by \cite{Parker83}. The majority of CSs formed probably do not lead to reconnection and "nanoflares". \cite{Einaudi21} recently pointed out that small scale temporally and spatially isolated CSs form the "elementary events", in their terminology, which play a key role in coronal heating. 

Recent observations using data taken by the Extreme Ultraviolet Imager (EUI) \citep{2020A&A...642A...1M} onboard the Solar Orbiter
mission \citep{2020A&A...642A...1M} show localized brightenings, termed ``campfires'', in quiet Sun regions. The identified events are mostly coronal in nature, with length scales between
400 km and 4000 km, and they seem to undergo internal heating all the way up to coronal temperatures \citep{Berghmans:5333}. The "campfires"  represent the small fraction of the observed reconnecting CSs \citep{ChenY21} in a "sea" of non-reconnecting current sheets (elementary events according to \cite{Einaudi21}), which act collectively, as we will show in this article,  on the particles of the ambient plasma and mainly heat them and accelerate a small fraction of ambient particles.

\cite{Fermi49} proposed a stochastic process to interpret the acceleration of cosmic ray particles, assuming that particles execute a random walk and elastically collide with 
``magnetic scatterers" and gain or lose energy. A few years after, Fermi returned with a new suggestion, introducing the turbulent shock as a mechanism where particles systematically gain energy \citep{Fermi54}. \cite{Sioulas20} used the methodology developed by Fermi to address the acceleration of particles inside a turbulent acceleration volume, using Monte Carlo simulations of particles inside a uniform or a fractal distribution of scatterers. The energy gain or loss of the particles interacting with a ``scatterer" depends on the physics of their local interaction.  In a large-scale turbulent reconnecting volume the magnetic cloud can be replaced by a CS or  "turbulent small scale structures" or both \citep{Vlahos04, Vlahos16, Pisokas16, Isliker17a, Pisokas18, Sioulas20, Sioulas20b}. 

The formation of a power law in energy space depends on the rate of the energy gain per ``collision" of a particle with a scatterer. The flow equation in energy space is 
\begin{equation}
    \frac{\partial{p(W,t)}}{\partial t} + \frac{\partial}{\partial W}\left(\left\langle \frac{dW}{dt} \right\rangle p(W,t)\right )=-\frac{p(W,t)}{t_{esc}},
\end{equation}

where $p{(W,t)}$ is the energy distribution function, $W$ the energy, and $t_{esc}$ provides an estimate of the rate at which particles are lost from the acceleration volume \citep{Longair11}. The escape time strongly depends on the transport of particles inside the acceleration volume. Assuming that for the high energy particles the rate of energy gain as a result of the interaction of the particles with the scatterers is given by the relation 
\begin{equation}
    \left\langle \frac{dW}{dt} \right\rangle =\frac{W}{t_{acc}},
\end{equation}
where $t_{acc}$ is the acceleration time, the steady-state solution of the energy flow equation is
\begin{equation}
     p(W)  \sim W^{-k},
     \label{eq:pWFermi}
\end{equation}
where $k=1+t_{acc}/t_{esc}$.

Energy release in large-scale astrophysical systems is a multi-scale process that couples the large-scale magnetic structure, where current fragmentation dominates in a 3D environment, with the small scale interaction of the particles with the coherent structures, e.g.\ CSs.
Understanding the mechanism for the energy release in the solar atmosphere requires a hybrid model that incorporates both the global evolution of the magnetic field and self-consistent feedback of the heating and acceleration of particles. We are far from reaching this stage yet.  

Numerical simulations of isolated periodic reconnecting current sheets have clearly demonstrated that particles are systematically accelerated  (see Fig.\ 3b in \cite{Li19} or Fig.\ 3b in \cite{Arnold21}). The heating of electrons in Particle In Cell simulations by reconnecting isolated current sheet has not been reported. On the contrary, turbulent reconnection initiated by a spectrum of MHD waves \citep{ComissoL19}, or by the fragmentation of a large scale current sheet formed through emerging magnetic flux \citep{Isliker19} present systematic acceleration of particles as they cross the reconnecting current sheets (see Fig.\ 9a in \cite{ComissoL19}  and Fig.\ 10d in \cite{Isliker19}), as well as stochastic heating of the low energy particles as they interact with coherent structures or non-reconnecting current sheets, as we call them in this article (see e.g.\ Fig.\ 11 in \cite{Isliker19} as an example of impulsive (collisionless) heating).

In this article, we assume that the turbulent reconnecting volume is a mixture of reconnecting CSs and non-reconnecting CSs.  We assume, based on the current literature mentioned above, that the interaction of the particles with the reconnecting CSs is systematic since the particles ``colliding'' with reconnecting CS undergo a very complex local interaction
\citep{Li19}. The interaction of the particles with the non-reconnecting current sheets \cite{Phan20} on the other hand is stochastic. We will show that the mixture of the two types of current sheets creates energy distributions with a hot core and a power-law-shaped high-energy tail during solar flares \citep{Lin03, ChenY21}.

In section 2, we present the model we are using to simulate a turbulent reconnection volume. In section 3 we apply our model to environments dominated by reconnecting and non-reconnecting CSs, and we analyze the synergy of a mixture of reconnecting and non-reconnecting CSs. In section 4 we summarize our main results.

\section{Initial set-up}
\label{sec:headings}

\begin{figure}[!ht]
     \begin{center}
         \includegraphics[width=1\textwidth]{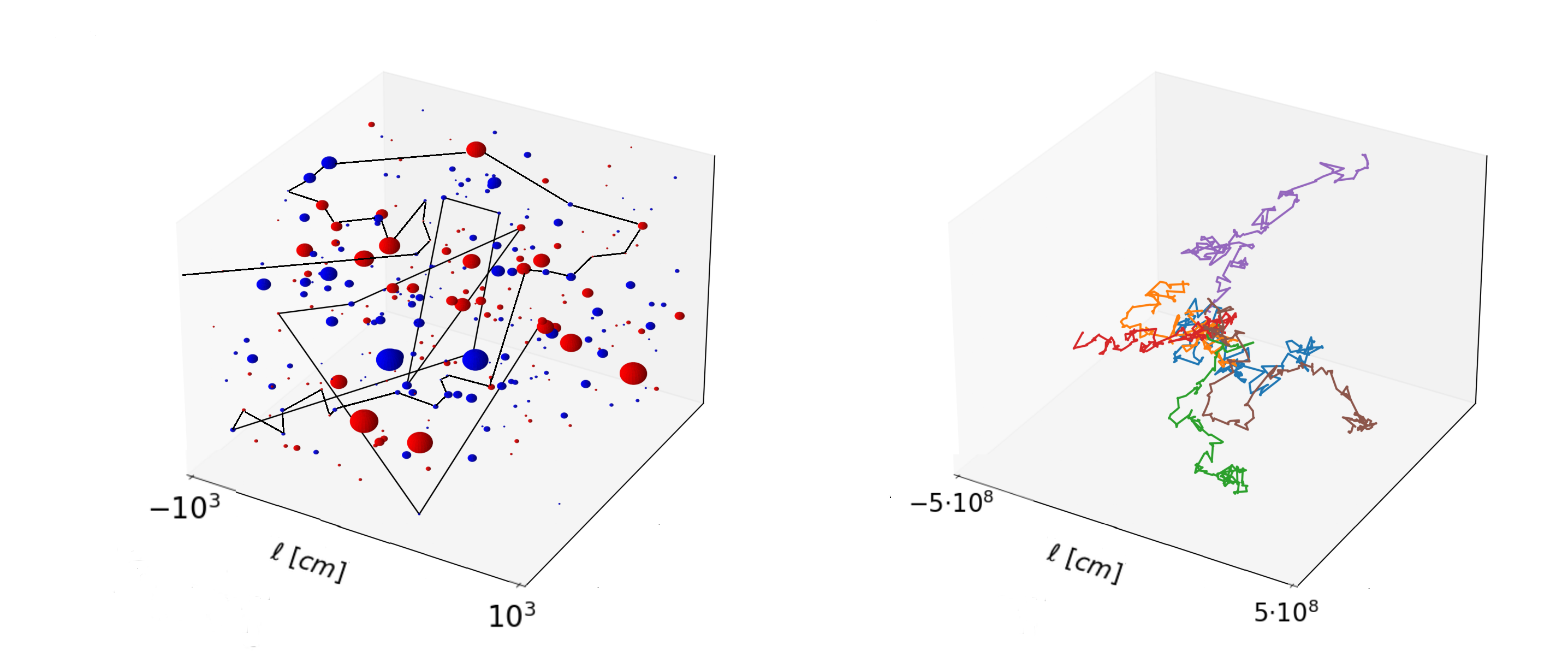}        
         \caption { (a) A schematic representation of the coherent current structures (CS) inside a zoomed view into the turbulent volume. The current filamentation creates a mixture of reconnecting (red) and non reconnecting (blue)  current sheets.  A particle executes a fractal random walk inside the turbulent volume. (b) Typical trajectories from our simulations for several particles, marked with different colors. The particles interact with an ensemble of CSs that form a fractal with dimension $D_F=1.8$ before they escape from the acceleration volume.
}\label{trajectory}
\end{center}
\end{figure}

We use a Monte Carlo code for the simulation of the turbulent reconnection environment, considering a three-dimensional box of linear size $L=10^{9}\,$cm. At time $t=0$ a population of $10^{6}$ particles is uniformly injected in the turbulent volume. Each particle is randomly assigned an initial energy $W_0$, such that the resulting distribution is a Maxwellian of temperature $T=100 eV$. Also, at $t=0$ all the particles are assumed to find themselves in the vicinity of a scatterer, which they immediately enter and undergo a first acceleration event. The particles are then allowed to move freely inside the acceleration volume until they once again meet a scatterer (see Fig. \ref{trajectory}a).

 The latter being distributed in such a way that they form a fractal set. \cite{Isliker03} analyzed this kind of
random walk where particles move in a volume in which a
fractal resides, usually traveling freely but being scattered when they encounter a part of the fractal set. They showed that in this case the distribution of free travel distances
$\lambda_{sc}$ in between two subsequent encounters with the fractal is distributed in good approximation according to 
\begin{equation}
    p(\lambda_{sc}) \sim \lambda_{sc}^{D_{f}-3},
\end{equation}
as long as $D_{F}   <  2$. For $D_{F}  > 2$, $p(\lambda_{sc})$ is decaying exponentially. Given that a dimension $D_{F}$ below 2 has been reported (\cite{Isliker03}, \cite{PhysRevE.65.046125}, \cite{Isliker19}), we are led to assume that $p(\lambda_{sc})$ is of
power-law form, with index between -1 and -3.
In \cite{Sioulas20b}, a parametric study of  ${\lambda_{sc}}_{min},\,{\lambda_{sc}}_{max}$ has been performed, which indicates that the results are not sensitive to the precise values of the limits as long as the difference between the scales is several orders of magnitude. We were thus led to assume that the spatial separation of the scatterers  ranges from ${\lambda_{sc}}_{min}=10^{2}\,$cm to ${\lambda_{sc}}_{max}=10^{9}\,$cm. The lower boundary represents the smaller-scale structures (i.e.\ $1\ {-}\ 10$ meters), while the higher limit is the characteristic scale of the acceleration volume.  For the value of the fractal dimension we follow \cite{Isliker19},  using $D_{F} =   1.8$, thus $ \alpha = 1.2$. Varying the fractal dimension in the range $D_{F} \ = \  1.8 \ \pm 0.2$ has minimal effects for the steady state energy distributions.

As a result, the series of distances $dr_{i}^{j}$, where $i =1,2,...,10^6$ is the particle index and $j=1,2,...,N_{i}$ the number of acceleration events for each particle, generated according to the probability density $P(dr)$, characterizes the trajectory of the $i$th particle in space. However, since the particles travel in three dimensional space, we also have to generate a random number for the azimuthal angle ${\phi}$, $0 <{\phi}< 2{\pi} $, and one for the polar angle $\theta$ through  $\cos{\theta}$, $-1 <\cos{\theta}< 1 $, which determine the direction of the particle motion. We can, therefore, monitor the coordinates of each particle, at the time of an energization event, according to

 \begin{align*}
x_{i}^{(j)}&=x_{i}^{(j-1)} + dr_{i}^{(j-1)}\cos{\phi} \sin{\theta}\\
 y_{i}^{(j)}
 &=y_{i}^{(j-1)} + dr_{i}^{(j-1)}\sin{\phi}\sin{\theta}\\
 z_{i}^{(j)}&= z_{i}^{(j-1)} + dr_{i}^{(j-1)}\cos{\theta}
 \end{align*}

After a scattering event, which we will here consider to be instantaneous, the particle's energy changes by the amount of ${\delta}W$ that is determined by the nature of the scatterer encountered. As a result, after the interaction, the particle $i$ performs it's $j$th spatial step of length $dr_{i}^{(j)}$ while having a constant velocity $v_{i}^{(j)}$
until the next encounter with a scatterer. Therefore, the time passed can be estimated by the free travel times. After a number of $N_{i}$ encounters, the total time elapsed for the $i$th particle is ${\tau_{i}^{N_{i}}}= \sum_{j=1}^{N_{i}}dr_{i}^{(j)}/|v_{i}^{(j)}|.$ We continue to monitor the particles' energy and transport properties up to the final simulation time  or until it crosses one of the box edges at time $t=t_{i,esc}$. Typical random trajectories of particles from our simulations are shown in Fig.\  \ref{trajectory}b. In order to keep track of the particles' energy and spatial evolution, we monitor their properties at a number of \textbf{logarithmically} equi-spaced monitoring times $t_{k}$, $k= 0,1,..., K$, with $K$  typically chosen as $10^{3}$ (see details in \cite{Sioulas20, Sioulas20b}).

The environment that we model is similar to the one present in the lower solar corona. As a result, we impose a magnetic field of strength $B = 100\,$G, an ambient temperature of $T=100\,$eV, and a plasma density of $n_{0} = 10^{9}\,$ cm$^{-3}$. Finally, the Alfven speed $V_{A}$
is analogous to the electron thermal speed, attaining a value $V_{A}\sim   7 \times 10^{8}\,$cm/sec. 

We focus our study on the random walk of ions and electrons in three different types of scatterers (1) reconnecting current sheets (section 3.1), (2) non reconnecting current sheets (Section 3.2), and (3) synergy of reconnecting and non reconnecting current sheets (Section 3.3). The processes outlined here can be applied in several laboratory and astrophysical settings where strong turbulence and turbulent reconnection are driven.

\section{Results}

\subsection{Systematic acceleration by Reconnecting Current Sheets ($RCS$s). }\label{UCS}



     

\begin{figure}[!ht]
  \begin{center}
          \includegraphics[width=1\textwidth]{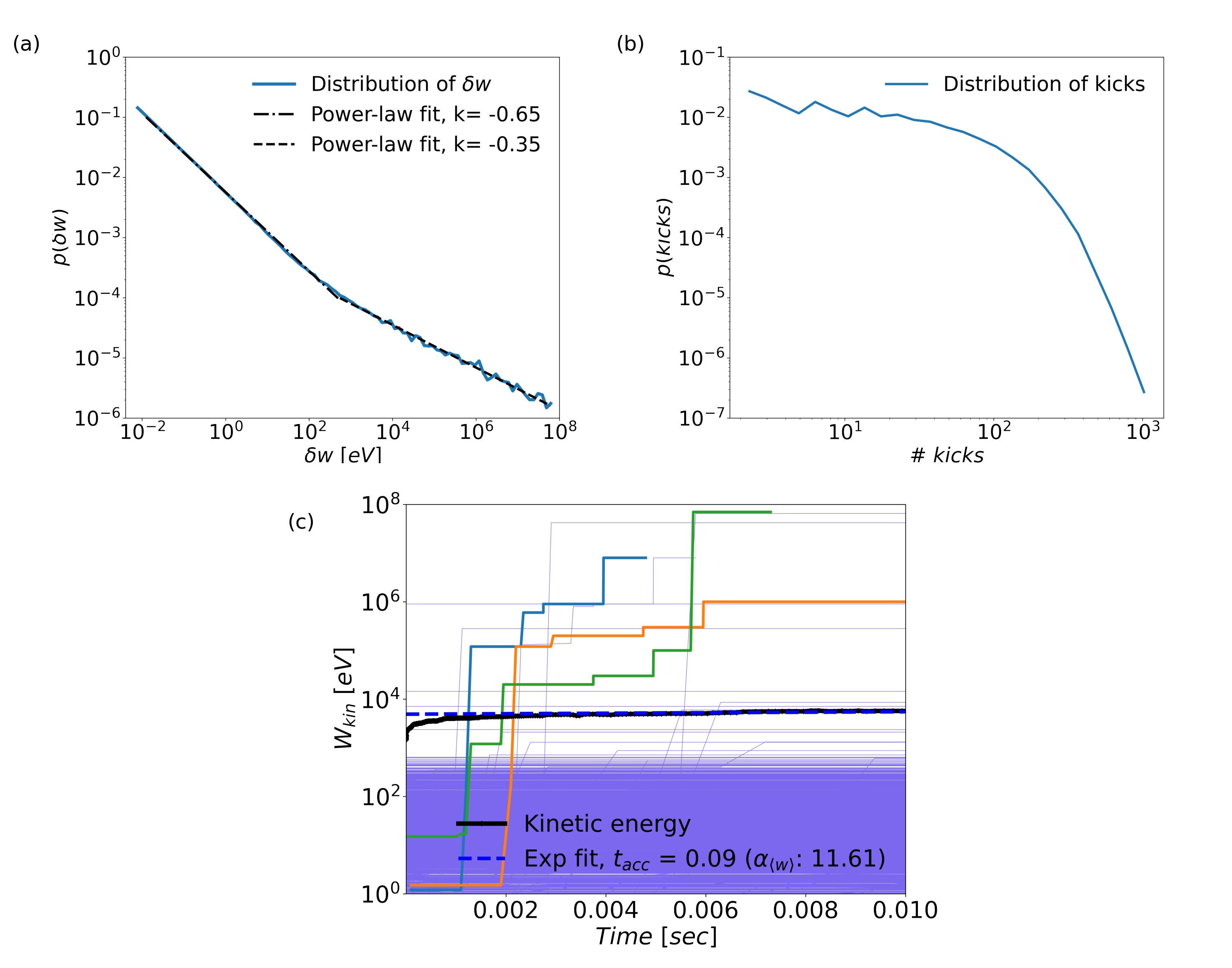}
        \caption {(a) The distribution of energy increments ${\delta}W$ for electrons interacting with  RCSs.  (b) The distribution of the number of energization events per electron.  (c) The mean kinetic energy as a function of time, along with some typical electron trajectories marked with different colors. }\label{fig2_new}
     \end{center}
     \end{figure}

  We begin our analysis by considering an environment where all of the scatterers are modelled to behave as  reconnecting current sheets (RCSs), systematically accelerating charged particles that travel in the turbulent volume. The energy gained after an interaction with a RCS is 
\begin{equation} \label{EnergygainUCS}
    \delta W=|q|(\frac{V_{A}}{c}\delta B){\ell}_{eff}
\end{equation}
\citep{Isliker17}. In our model, the magnetic fluctuation $\delta B$ takes random values in the range ${\delta} B \in [ 10^{-5} G, 10^{2} G$],  obeying the probability density $P(\delta B)  \sim  \delta B^{-5/3}$ \citep{Pisokas18}. As a result, the effective electric field $E_{eff} =(\frac{V_{A}}{c}) \delta  B $ takes values in the range $E_{eff} \in [ 2 \cdot 10^{- 7}\, \mathrm{statV/cm},2\, \mathrm{statV/cm} ]$. Finally, we consider the effective length (${\ell}_{eff}$) to be a linear
function of $E_{eff}$, ${\ell}_{eff} = aE_{eff} + b$, where the constants a, b can be estimated by limiting ${\ell}_{eff}$ to the range  ${\ell}_{eff} \in [1 m,1 km]$. If we apply these values to
Eq.\ \ref{EnergygainUCS}, we can estimate that $\delta W$ varies between $6{\cdot}10 ^{- 3} eV$ and $10^8 eV$ and follows a double power-law, see Fig.\ \ref{fig2_new}a  \citep{Isliker17, Pisokas18}. In energy space, the particle dynamics thus exhibit a systematic random walk, as only positive increments occur. In  Fig.\ {\ref{fig2_new}}b, the distribution of the total number of acceleration events per particle is presented. The number of kicks ranges from 2 to 1055, with a mean of $ \sim 72$ acceleration events per particle. Following a scattering event, the $i$th particle departs from the scatterer with its renewed energy, 
$$W_i^{j+1}=W_i^j+\delta  W_i^j .$$
Here, $\delta W_i^j$ is given by Eq.\ \ref{EnergygainUCS}, and $j$ counts the number of energization events for the particle, up to a given time. 

\begin{figure}[!ht]
     \begin{center}
    
       \includegraphics[width=1\textwidth]{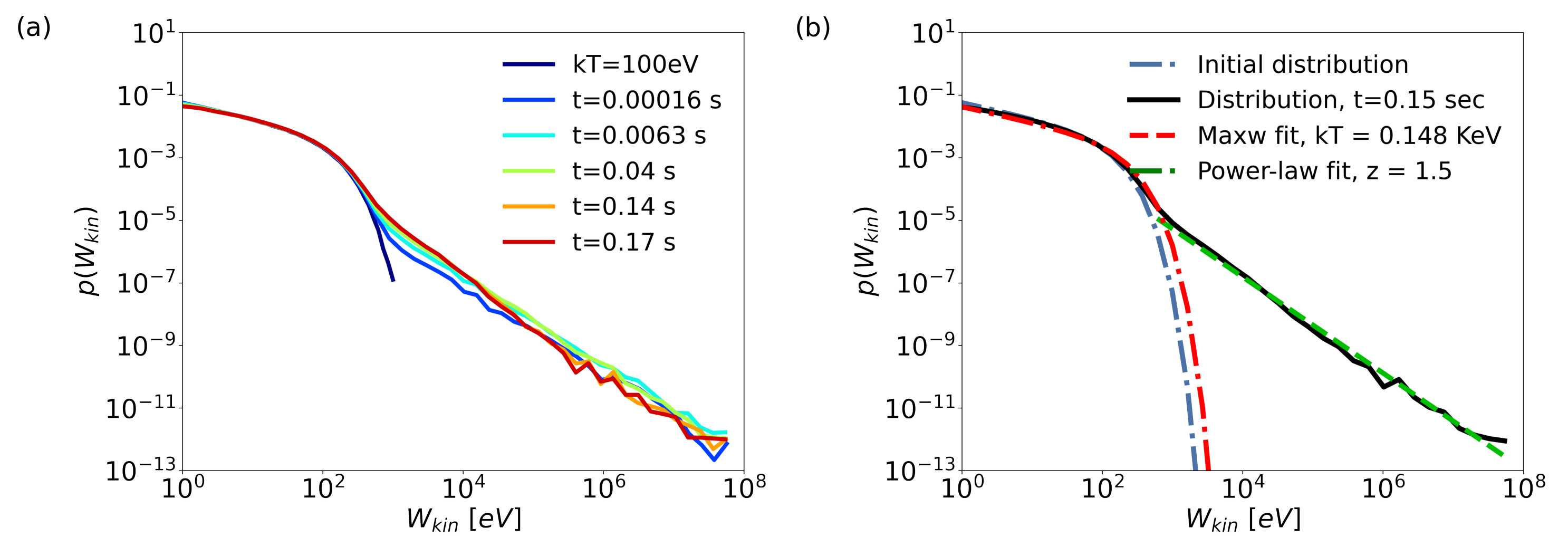}
       
     \caption {(a) Temporal evolution of the kinetic energy distribution of the electrons in the presence of RCSs.  (b) Kinetic energy distribution at $t = 0$ and $t = 0.15\,$s (steady state) for the electrons remaining inside the volume with size $L=10^{9}\,$cm, together with  a Maxwellian fit at low energies and a power-law fit at high energies.}\label{evol}
     \end{center}
     \end{figure}

  In Fig. \ref{fig2_new}c, we plot the evolution of the mean energy $<W(t)>$ of the particles and the energy gain of a few typical particles along their trajectories. The mean energy initially increases  exponentially in time, $<W(t)> \sim e^{a_W t}$, and the acceleration time can be estimated from the relation $t_{acc} =1/a_w \sim 0.09 sec$. The bulk of the particles gains only an almost negligible amount of energy from the interactions with the reconnecting Current Sheets. Only a small fraction of particles gains systematically energy, forming a  power-law tail for energies $W_{kin}>1\,$keV, see Fig.\ \ref{evol}.   The interactions of particles with the scatterers pull particles from the thermal part of the distribution to super{-}thermal energies, resulting in a persisting tail (Fig. \ref{evol}a). The bulk of the distribution basically remains unaffected and essentially is not heated by the presence of reconnecting current sheets.

\begin{figure}[!ht]
     \begin{center}
      \includegraphics[width=1\textwidth]{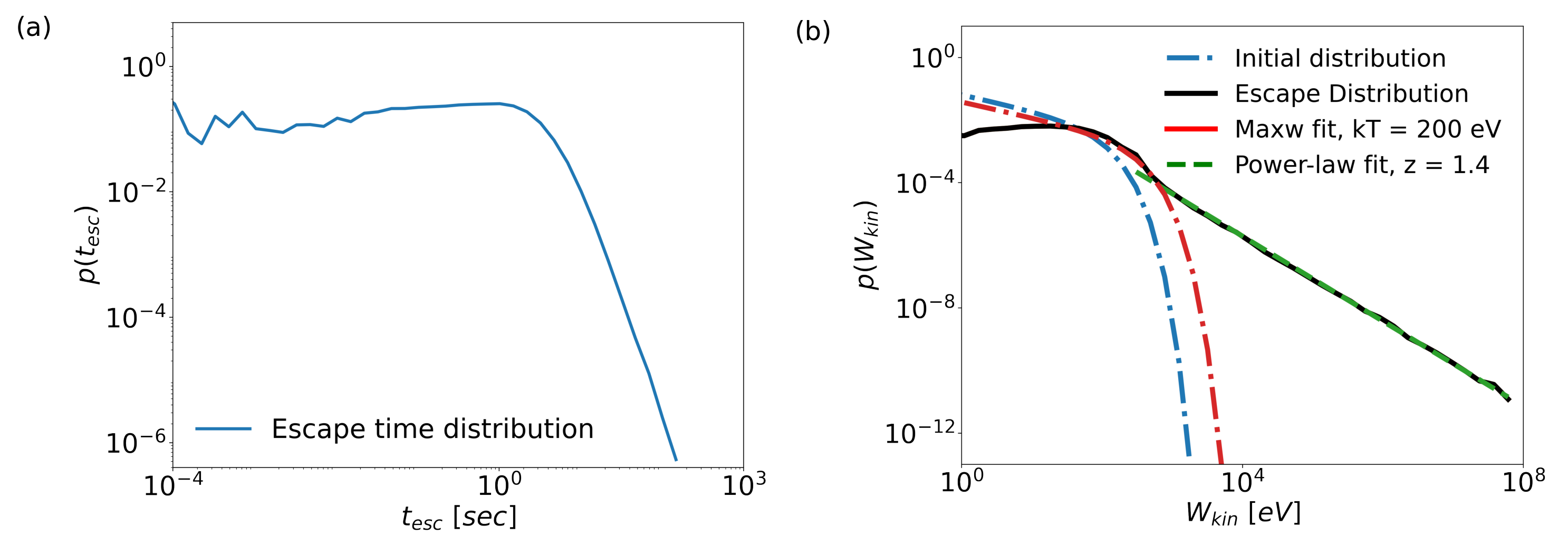}
       \caption {(a) The escape time distribution.  (b) The  escape energy distribution of the electrons. }\label{escape}
        \end{center}
     \end{figure}

The evolution of the kinetic energy distribution in Fig.\   \ref{evol}a indicates that the high{-}energy tail forms in a matter of a few milliseconds and persists even when more electrons have escaped from the volume, which in turn implies a very fast acceleration process. The distribution reaches its asymptotic state after a period of $t\ {\sim}\ 0.15$ sec, and the steady-state power{-}law tail yields an index $k\ =\ 1.5,$ which is close to $k=1+t_{acc}/t_{esc}{\sim}1.5$, see Eq.\ \ref{eq:pWFermi}, where $t_{esc} \ = \ 0.18 \sec$, the median value of the times at which the electrons depart from the simulation volume.

The distribution of the particles' escape time is shown in Fig.\ \ref{escape}a.  Taking into account the energy at which a particle leaves the box, denoted as $W_{esc}$, we show in Fig.\ \ref{escape}b the escape energy distribution, which follows a scaling similar to that of the distribution of particles remaining inside the volume. Studying the escape time as a function of the escape energy of the particles, we conclude that the super{-}thermal particles (i.e. $W_{esc}\ {>}\ 10^{3}\ eV$ ) tend to depart faster from the box in comparison to the low energy ones.



In the following, the role of the simulation box's length, as well as the magnetic fluctuations' strength ($\delta$B) is examined. Keeping all the other parameters constant, we reduce the maximum value of the magnetic fluctuations' strength to ${\delta}B \ = \ 10 \ G$. As a result, ${\delta}W$ takes values in the range $[7{\cdot}10^{-3}\ eV, 7{\cdot}10^{6}\ eV]$. It turns out that the steady-state energy distribution is similar to the one in Fig.\ \ref{evol}b, with the only difference being that the power-law tail is shorter and extends to a value that coincides with the maximum of ${\delta}W \ {\sim} \ 7{\cdot}10^{6}\ eV$. This behavior (i.e.\ the distribution to preserve its shape with a tail that extends up to ${\delta}W_{max}$ ) seems to hold in any of the cases tested. Also, other important parameters such as the escape and acceleration time do not seem to be considerably affected when reducing ${\delta}B$.
     Finally, in the case of RCSs, reducing the size of the simulation box, to still logical values, does not affect the shape of the steady-state energy distribution or the index of the power-law tail. As shown in Fig.\ \ref{fig2_new}, the probability density $P({\delta}W)$ follows a power{-}law scaling and extends over many orders of magnitude. Although the probability for a very efficient energization event that will result in a significant amount of energy gain is low,  we always have a few particles accelerated to those high energies, even at the very first stages of the acceleration (i.e.\ $t\ < 1 \ ms $). As a result, the high-energy tail is always present.  One other important observation is that there is a strong correlation between a significant energization event and the immediate departure of a particle from the acceleration volume. 

\begin{figure}[!ht]
     \begin{center}
       \includegraphics[width=1\textwidth]{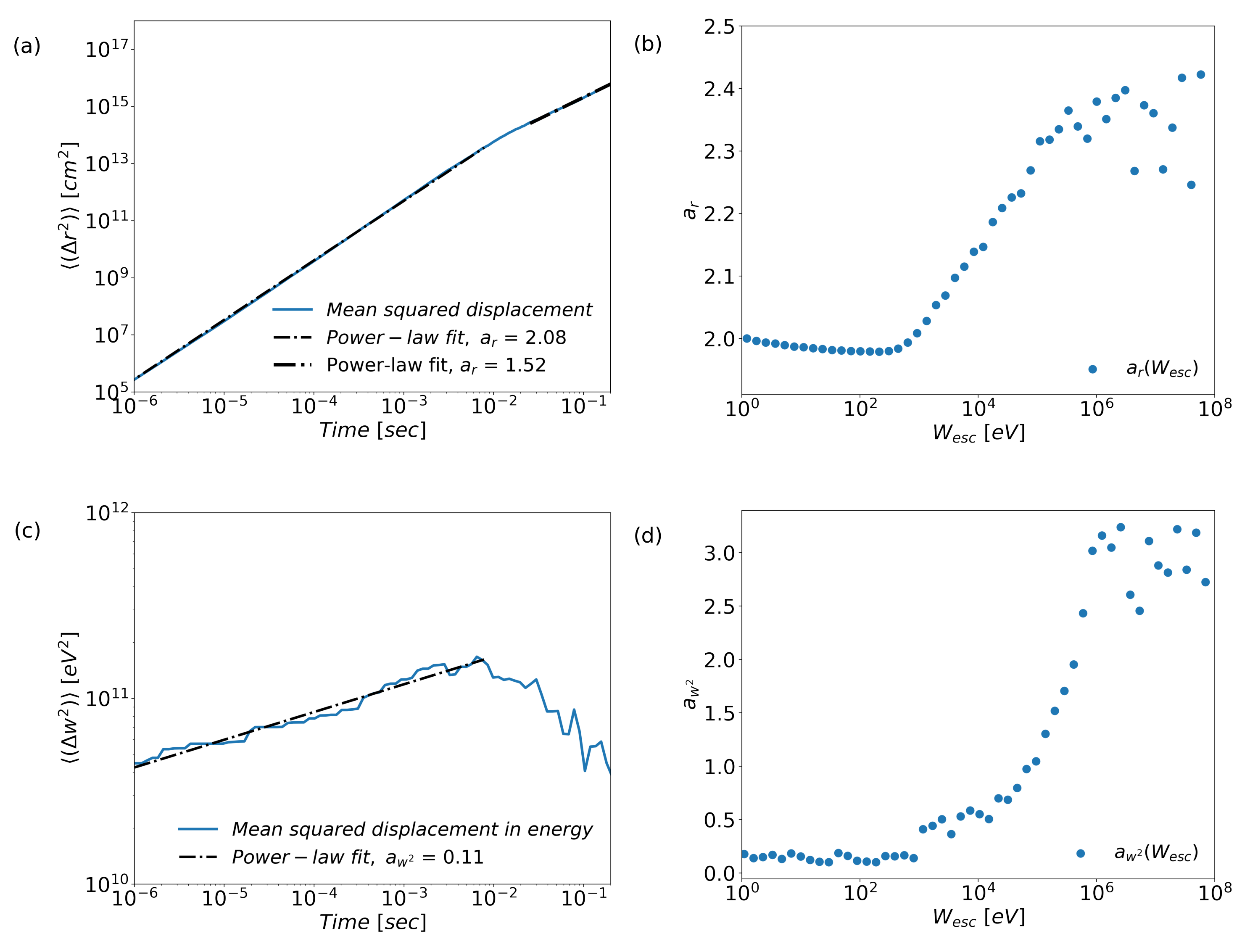}

         \caption {(a) Spatial mean squared displacement as a function of time for the electrons interacting with RCSs. (b)  Scaling index of the spatial mean squared displacement as a function of the escape energy. (c) Mean squared displacement in energy as a function of time for the electrons. (d) Scaling index of the mean squared displacement in energy as a function of the escape energy.
}\label{rmsd1}
     \end{center}
     \end{figure}

Using the set of predefined monitoring times described above, we continue to keep track of the particle population properties up to the point where the majority of them has escaped, namely at time $t \ {=}\ 1$ second.
 A particle's displacement from its initial position at time $t\ {=}\ t^{n}$  is ${{\Delta}\vec{r}_{i}^{n}}=\vec{r}^{n}_{i}-\vec{r}_{0i}$, and the mean square displacement for the ensemble of particles can be estimated through
 \begin{equation} 
     {\langle}({\Delta}r^{n})^{2}{\rangle}=\frac{1}{N_{p}}\sum_{i=1}^{N_{p}}({\Delta}r^{n}_{i})^{2}. \label{msqd} 
 \end{equation}
 
 To test the validity of the test-particle code, we first simulate an environment, in which a scattering event exclusively changes a particle's direction of motion, leaving its energy constant (passive scatterers).   In this case, we find that the diffusion of the charged particles is ballistic, with the power-law index holding a value of $\sim 2$. This result is in good accordance with the results obtained by \cite{Isliker03} (as shown in Figs.\ 10 and \ 11 therein), where also the particles interact with passive scatterers that reside on a fractal with dimension $D_{F}<2$. Turning to active scatterers, the mean squared displacement is presented in Fig.\ \ref{rmsd1}a as a function of time. As one can observe, the charged particle diffusion is a two-stage process. For times up to $t = 10^{-2} sec$ we have  an impulsive phase, which  exhibits a ballistic scaling $<(\Delta r)^2> \sim t^{2.08}$. At larger times, the high-energy particles have departed from the volume and do not contribute any more to the statistics, which in turn leads to a reduction of the degree of super-diffusivity. As a result, for times greater than $t = 10^{-2}$ sec the power-law index decreases to ${\alpha_r} = 1.52$.  In Fig.\ \ref{rmsd1}b, we have divided the particles into 50 logarithmically equi-spaced bins, according to the energy with which they escape the turbulent volume $W_{esc}$.  
As expected, the particles can be divided into three populations. The thermal particles ($W_{esc} \ {\leq} \ 10^{3} \ eV$) that with respect to their spatial diffusion properties behave similarly to particles interacting with passive scatterers, and the super-thermal particles, for which the spatial diffusion process is more efficient, with the power-law index linearly increasing from 2 to 2.45 and then saturating as the escape energy attains larger values.

The mean square displacement  in energy can be estimated through
    \begin{equation}\label{sqenergytr}
    \langle (\Delta  W)^2\rangle (t^n) \equiv \langle (\Delta  W^n)^2\rangle =\frac{1}{N_p}\sum_{i=1}^{i=N_p} (\delta W_i^n)^2.
    \end{equation}


        
    
 In Fig.\ {\ref{msqd}}c,  we show the mean square displacement in energy as a function of time. It follows a power{-}law scaling of the form $ \langle (\Delta W)^2\rangle (t) = D_{W^2} t^{a_{W^2}}$ with index, ${\alpha}_{w^{2}}\ = \ 0.11$, indicating a sub-diffusive process in energy space. However, this result can be misleading. Similar to the spatial diffusion, if we divide the particles into logarithmically equi-spaced bins according to their escape energy, we find three distinct populations, 
 see Fig.\ {\ref{msqd}}d. The bulk of the distribution consists of the thermal particles that diffuse in energy space in a sub-diffusive manner, barely gaining energy before they escape from the simulation volume. On the other hand, a small fraction of the particles are accelerated to high energies ($W_{esc} \ \geq \ 10^{3} eV$), for which the energy transport process is very effective, as indicated by the power-law index that increases from 0.1 to 3.5 before it saturates for escape energies of very large values.

  \subsubsection{Ion Acceleration by RCS's.}

   \begin{figure}[!ht]
     \begin{center}
     
         \includegraphics[width=0.5\textwidth]{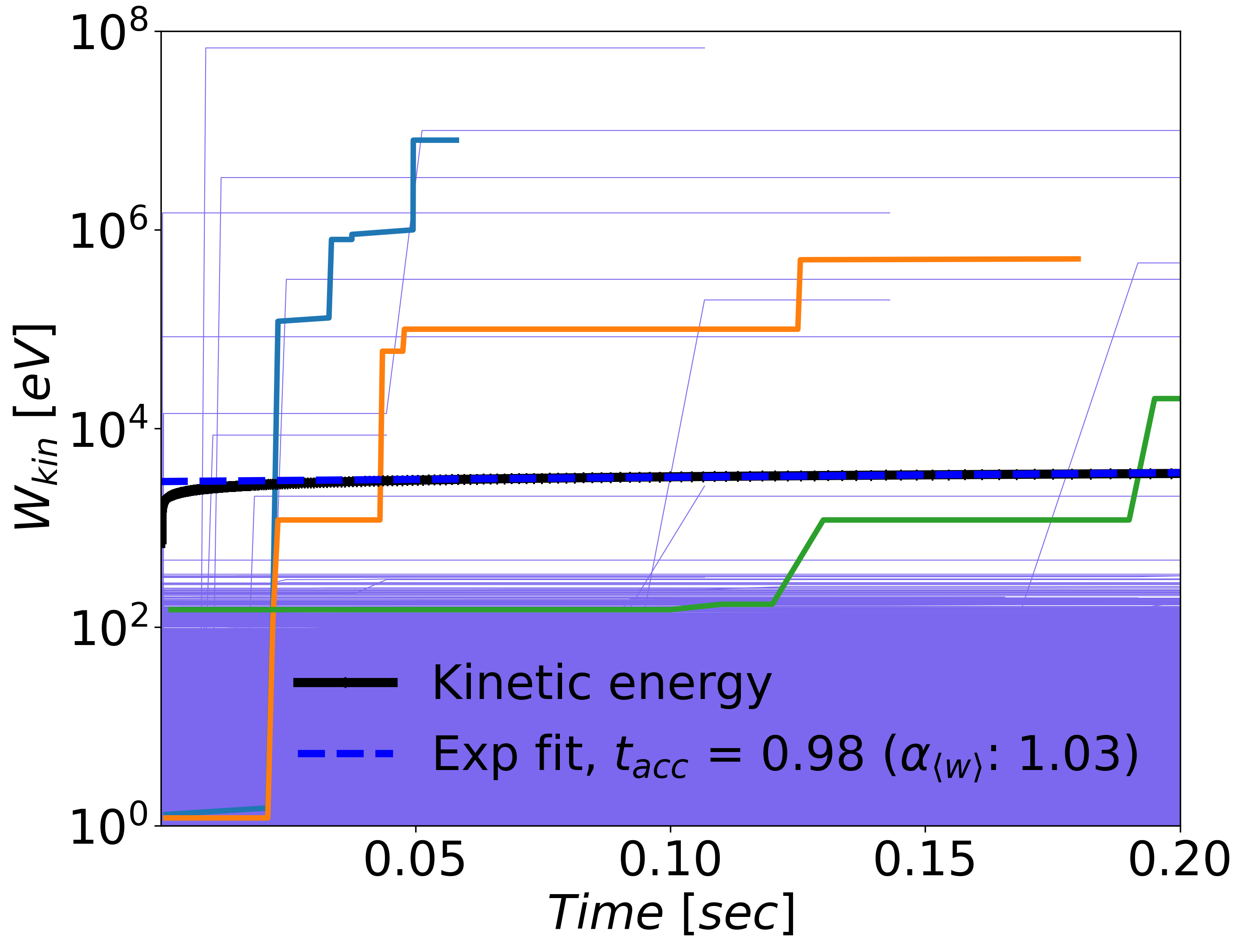}
    
  \caption {The mean kinetic energy as a function of time, along with some typical proton trajectories marked with different colors. }\label{Dw_ions_UCS}
  \end{center}
     \end{figure}

  Using the setup described in Sec.\ \ref{UCS}, we now replace the injected electron population with an equal number of ions (protons). Once again, the initial ensemble of the injected particles follows a Maxwellian distribution of temperature $T \ = \ 100 \ eV$. Since the energy increments ${\delta}W$ do not depend on the mass of the particles, their distribution is identical to the one for electrons (Fig.\ \ref{fig2_new}a)). What changes, in this case, is the time between subsequent energization events. An exponential fit to the mean kinetic energy curve in Fig.\ \ref{Dw_ions_UCS} yields $t_{acc}=0.98\,$s and shows that the energization process for protons is much slower than the one for electrons and is characterized by big time-gaps in-between scatterings with parts of the fractal. However, before escaping the acceleration volume, the protons are subjected on average to the same number of scattering events as the electrons.  
  
 
  \begin{figure}[!ht]
     \begin{center}
       \includegraphics[width=1\textwidth]{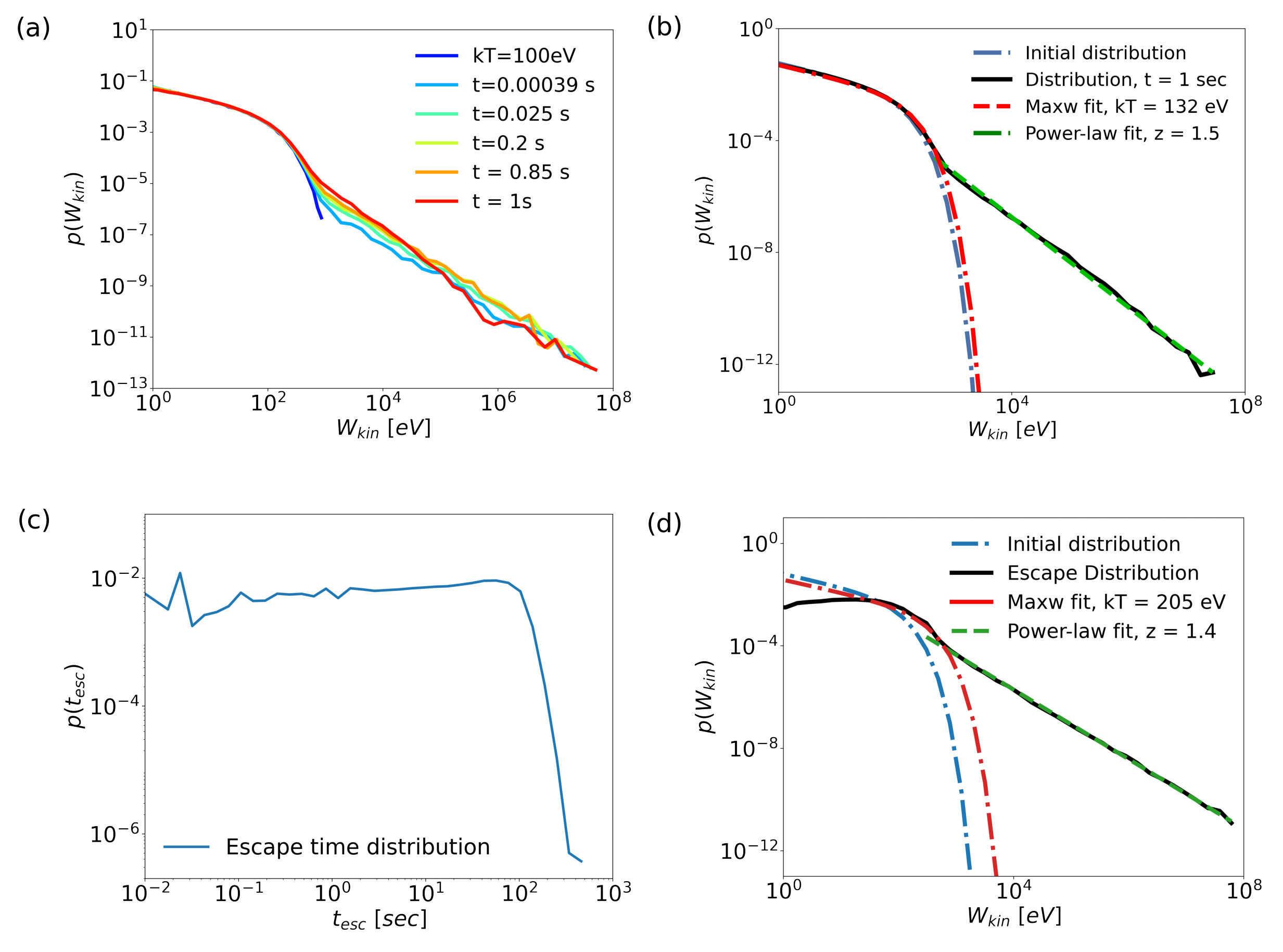}
     \caption {(a) Temporal evolution of the kinetic energy distribution for protons.  (b) Kinetic energy distribution at $t = 0$ and $t = 1\,$s (steady state) for the protons remaining inside the box with size $L=10^{9}\,$cm, together with  a Maxwellian fit at low energies and a power-law fit at high energies. (c) Escape time distribution for the protons. (d) Escape energy distribution for the protons, along with a power-law fit and a Maxwellian fit yielding a temperature $T \ = \ 205 eV$. 
     }\label{evol_ions_UCS}
     \end{center}
     \end{figure}

  As shown in Fig.\ \ref{evol_ions_UCS}a, the power-law tail of the kinetic energy distribution is formed almost instantly (in a matter of a few milliseconds). However, it takes ${\sim} \ 1 \sec$ for the energy distribution to reach its steady state, which exhibits the same characteristics (i.e.\ minor heating accompanied by significant particle acceleration) as in the case of  electrons. At $t=1\,$s (Fig.\ \ref{evol_ions_UCS}b), the lower energy part of the distribution can be fitted by a Maxwellian of temperature $T \ = \ 132 \ eV$, while for higher energies ($W_{kinet} \ {\geq} \ 10^{3}\ eV$) a power-law fit yields an index $z \ = \ 1.5$.

 In Fig.\ \ref{evol_ions_UCS}c, the distribution of the escape times for the protons is presented. The distribution remains uniform for escape times up to $t{\sim}10^2 \ sec $ and turns into a power-law with index $z\ = \ 7.2$ for larger escape times.  The median value of the escape times is  estimated as $t_{esc} {\sim} 21.8 \sec$ . Keeping track of the energy at which each particle has escaped the acceleration volume as $W_{esc}$, we present in Fig.\ \ref{evol_ions_UCS}d the escape energy distribution. The low energy part of the distribution can be fitted by a Maxwellian of temperature $T  =  205 eV$, while for escape energies greater than $10^{3}$ eV a power-law tail of index $z  =  1.4$ is formed. 

 Using binned statistics, we can determine the number of scatterings as a function of the escape energy.   For the thermal particles (i.e.\ $W_{esc} \ {\geq} \ 10^{2} \ eV$), the escape energy is strongly influenced by the number of scatterings a particle is subjected to during the energization process. For higher escape energies, the functional form becomes, in approximation, a constant, attaining a value of the order of $ {\sim}$ 100 scatterings per particle. The escape time is thus a function of the escape energy for the protons.

 The mean squared displacement  exhibits a power-law scaling ${\langle}({\Delta}r)^{2}{\rangle}{\sim} \ t^{2.06}$, which, after a time-span of $t \ {\sim} \ 0.1 \ sec$,  decreases to $t^{1.40}$, indicating a considerably less impulsive diffusion process than for electrons. This phase is also the longest-lasting one, as it persists for times up to $t  {\sim} 100 \ sec$, the time where the majority of the particles have departed from the acceleration volume. Just like in the electron case, the change in the spatial diffusion behavior happens at the same time where the mean kinetic energy reaches its maximum value. Comparing the protons' spatial diffusion to the electrons' one, we can see that protons require a considerable amount of extra time in order to spread and escape the simulation box. This is reflected in both stages of diffusion, with the spatial diffusion scaling being characterized by smaller values of the power-law index $a_{r}$. As a result, a significant increase in the escape time for the protons is observed when compared to the electrons. 
  

\subsection{Particle Heating by Non{-}Reconnecting Current Sheets ($CS$s). }\label{non-rec-UCS}

 The main differences between reconnecting and non{-}reconnecting current sheets are (1) an encounter with the former results in systematic (i.e.\ ${\delta} W$ is always positive ), while with the latter in stochastic energy change. (2) The absence of reconnection, which would otherwise strongly enhance the effective electric field $E_{eff}$, results in a considerable reduction in the amount of energy change ${\delta}W$ for the particles. Similar to the case of reconnecting current sheets, and according to the equation
 \begin{equation} \label{energygain_stochastic}
     W_i^{j+1}=W_i^{j+1} \pm \delta W_i^{j+1},
 \end{equation}
 the amount of energy change after an interaction with a non-reconnecting CS is assumed to be 
 \begin{equation}\label{energygain_stocastic_1}
     \delta W = |q|E_{eff} {\ell}_{eff}.
 \end{equation}
  Depending on the particle velocity direction relative to the direction of the effective electric field, the encounter with a non-reconnecting current sheet can either lead to energy gain or energy loss.
  
  As a result, in order to model this stochastic accelerator, the knowledge of two probability density functions is required:

       \begin{enumerate}
      
      \item The probability density $P(\ell_{eff})$ defines the effective acceleration lengths of the scatterers, ${{\ell}_{eff}}_{i}^{j}$. Here, ${\ell}_{eff}$ is assumed to follow a Kolmogorov spectrum, i.e.\  $P({\ell}_{eff})\ {\sim}\ {{\ell}_{eff}}^{-5/3}$, with the length of the scatterers taking values in the range ${\ell}_{eff} \in  [1 m,1 km]$,  \cite{Zhdankin13}.
      
      \item The probability density $P(E_{eff})$ yields the effective electric field $E_{eff,(i)}^{(j)}$ acting on the $i$th particle during its encounter with a non-reconnecting CS, after it has performed its $j$th spatial step. In this case, since no respective studies suggesting the exact form of this PDF exists, we were obliged to study a multitude of different distributions. 
      
           \end{enumerate}
      
      We will start our analysis assuming that $P(E_{eff})$  follows a narrow Gaussian distribution. As such,  the effective electric field is almost constant across all of the scatterers in the turbulent volume, and what varies is each scatterer's effective length. Another argument that satisfies this choice is its compatibility with the Central Limit Theorem (CLT). The effective electric field attached to each UCS can be thought of as the result of a plethora of small processes, a procedure that according to the CLT results in Gaussian distributions.  
      We consider for the purposes of this study a Gaussian of mean 
      ${\mu\ {=}\ 10^{3} {\cdot} E_{D}}$, and standard deviation  
      ${\sigma\ {=}\  10 {\cdot} E_{D}}$,
        where $E_{D}$ is the Dreicer field,
        $E_{D} \sim 1.6 \cdot 10^{-7}\,$statV /cm.

\begin{figure}[!ht]
     \begin{center} 
         \includegraphics[width=1\textwidth]{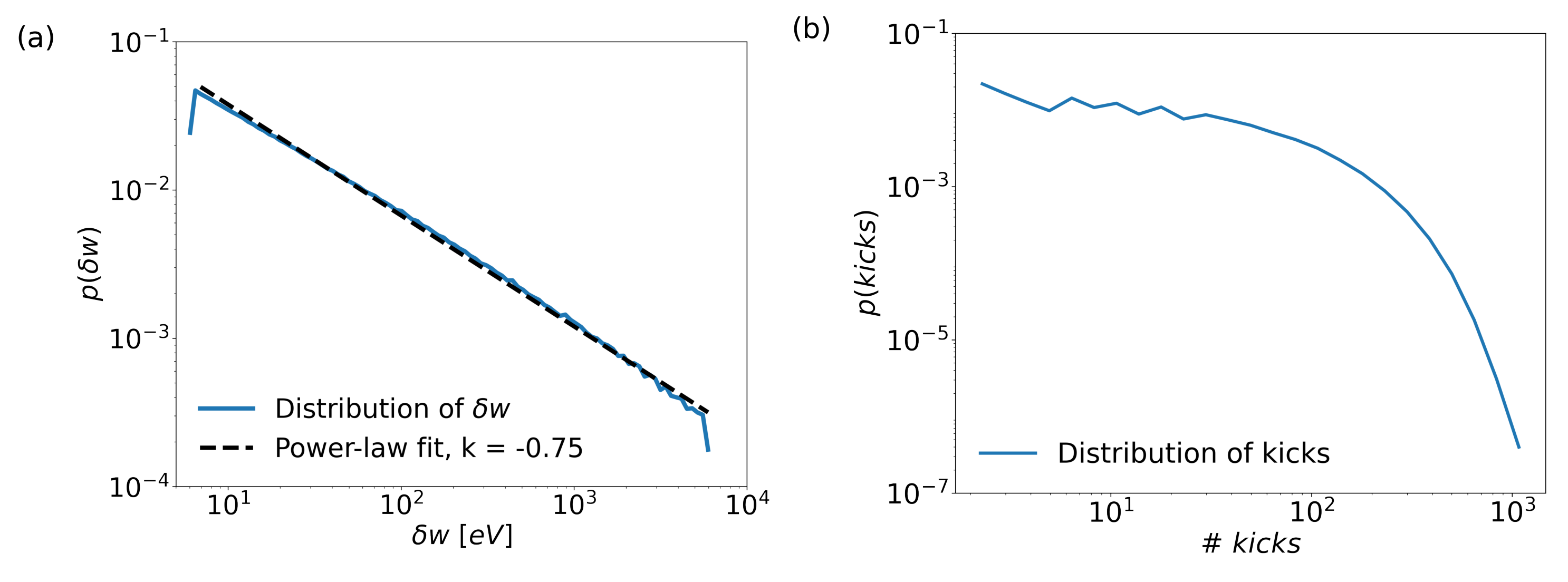}
         \caption {(a) The distribution of the absolute values of the increments $ \delta W$ resulting from the Eq.\ \ref{energygain_stocastic_1}, together with a power-law fit. (b) The distribution of the number of kicks per electron.}\label{fig8_new}
     \end{center} 
     \end{figure}

 As a result of the selected PDF's, the \textbf{absolute} values of $\delta W$ are ranging between $[3 \  eV, \ 8 \cdot 10^{3}\  eV ]$ with a median value ${\sim} 1.6 \cdot 10^{1} eV$, and their distribution forms a power-law, see Fig.\ \ref{fig8_new}a.
  
  \begin{figure}[!ht]
     \begin{center}
         \includegraphics[width=1\textwidth]{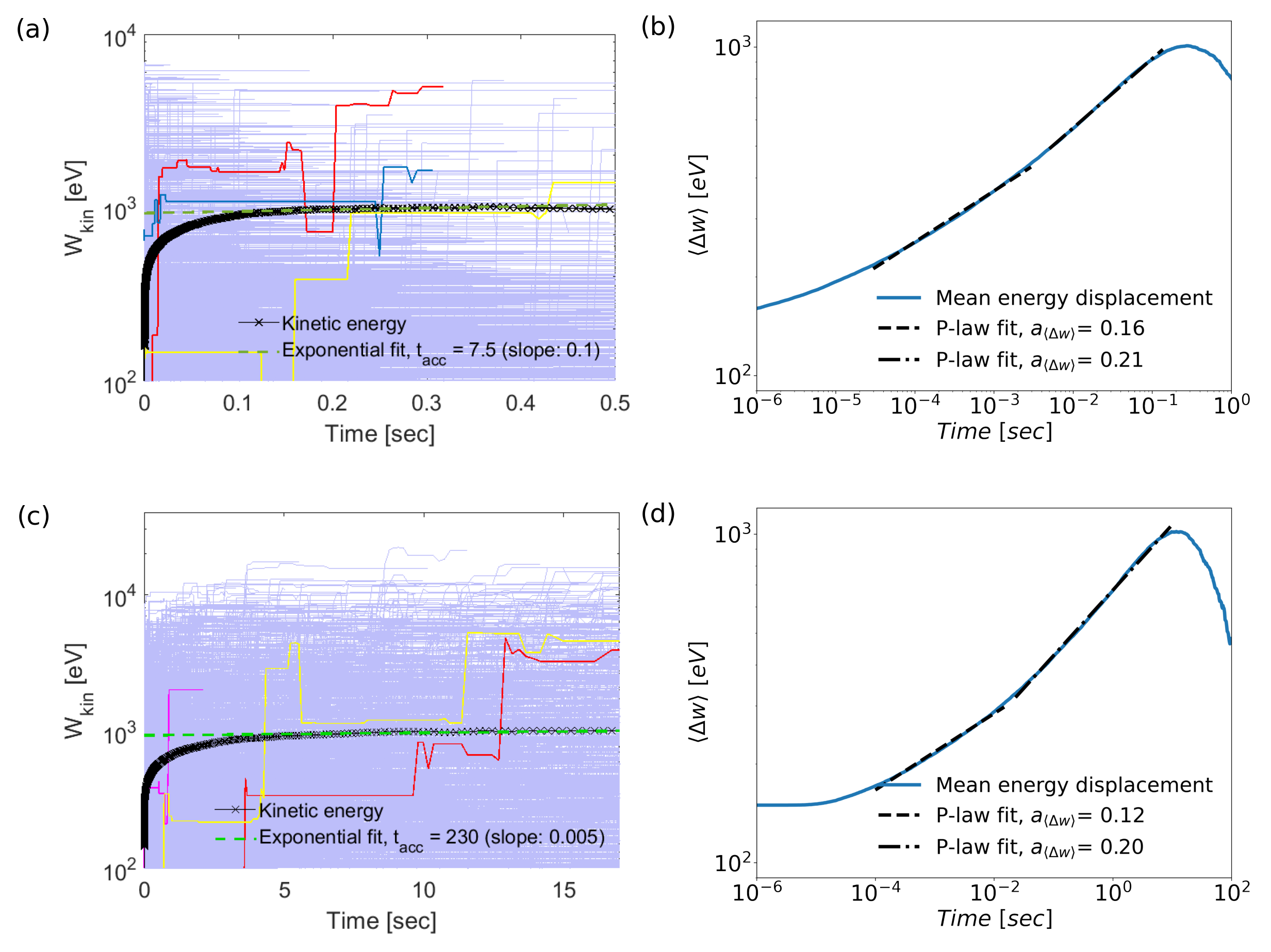}  
         \caption {(a) The mean kinetic energy as a function of time, and the energization process for several selected electrons. (b)  Mean  displacement in energy as a function of time for the electrons.  (c) The mean kinetic energy as a function of time, and the energization process for a number of selected ions. (d)  Mean displacement in energy as a function of time for the ions. }\label{Dw_non_rec}
  \end{center}
     \end{figure}

    We begin our analysis by uniformly inserting either a population of $10^{6}$ electrons, or an equal number of ions (protons) inside the turbulent domain. The particles in their ensemble follow a Maxwellian distribution of temperature $T \ = \ 100 eV$. Following the setup of predefined monitoring times presented in Sec.\ \ref{UCS}, we keep track of the transport properties of the energized particles. During the acceleration process, each particle is subjected to many acceleration events (kicks), before it escapes the simulation box at time $t\ = \ t_{esc,i}$.

     \begin{figure}[!ht]
     \begin{center}
         \includegraphics[width=1\textwidth]{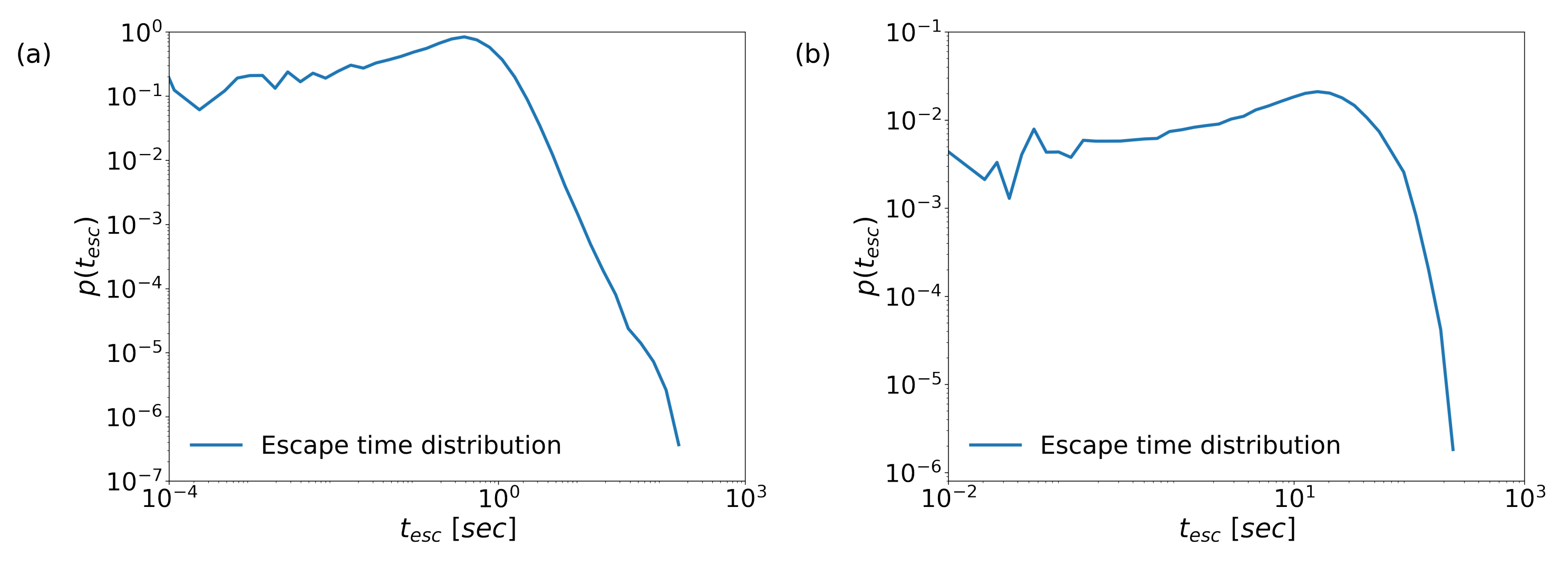}
         \caption {(a) The distribution of the electrons' escape time.  (b)  The distribution of the protons' escape time.
        }\label{esc_time_non_rec_UCS}
     \end{center}
     \end{figure}

    Fig.\ \ref{fig8_new}b exhibits the distribution of kicks per particle for the electrons. For the electron population, the mean value of the number of energization events is ${\sim}\ 94$, with the number of kicks per particle ranging from 2 to 1232.  The protons' distribution of kicks per particle is very similar to the one of the electrons. The number of energization events per ion also has a wide range from 2 to 1167,  with a mean value of ${\sim}\ 91$ events.
    Thus, in terms of acceleration events, the two particle populations behave in a very similar manner. 
    
    In Figs.\ \ref{Dw_non_rec}a,c, we present the energization process for a number of selected electrons and ions, respectively. Apart from the stochastic nature of the acceleration process, what also becomes apparent is the fact that although ions attain approximately the same energies as the electrons, they require some tens of seconds more time inside the acceleration volume to do so. Fig. \ref{Dw_non_rec}b,d present the mean displacement in energy as function of time. Applying an exponential fit to the mean kinetic energy, one can estimate the acceleration time for each species, t$_{acc}\ =  1/a_{w}$, which is a good indicator of the rapidness of the acceleration process. We in this way estimate t$_{acc}\ {\sim} \ 10 \sec $ for the electrons, and, t$_{acc}\ {\sim} \ 200 \sec$ for the protons. Another timescale that differs between the two populations is the one that refers to the time they spend in the acceleration volume, $t_{esc}$. In Figs.\ \ref{esc_time_non_rec_UCS}a,b, the distributions of the electron and proton escape times are presented. For the electrons, the distribution is in a good approximation uniform for times up to  $t {\sim} \ 1 \sec$ and turns into a power-law with index $z \ = \ 2.7$ for larger times. The proton escape time distribution is of similar shape, namely uniform for times up to ${\sim} \ 90 \sec$, followed by a steep power-law part with index $z \ = \ 8.1$.  The median value of the electron escape time is estimated as $t_{esc} \ = \ 0.7 \sec$, while for the protons as $t_{esc} \ = \ 26.9 \sec$. The median escape times thus indicate that in general electrons tend to depart much faster from the acceleration volume than protons. It is important to note that the exact values of the escape and acceleration time strongly depend on the mean value of the imposed Gaussian distribution of the effective electric field. More specifically, increasing the distribution's mean results in a reduction in both, the escape  $t_{esc}$ and the acceleration $t_{acc}$ time, respectively.

\begin{figure}[!ht]
     \begin{center}
         \includegraphics[width=1\textwidth]{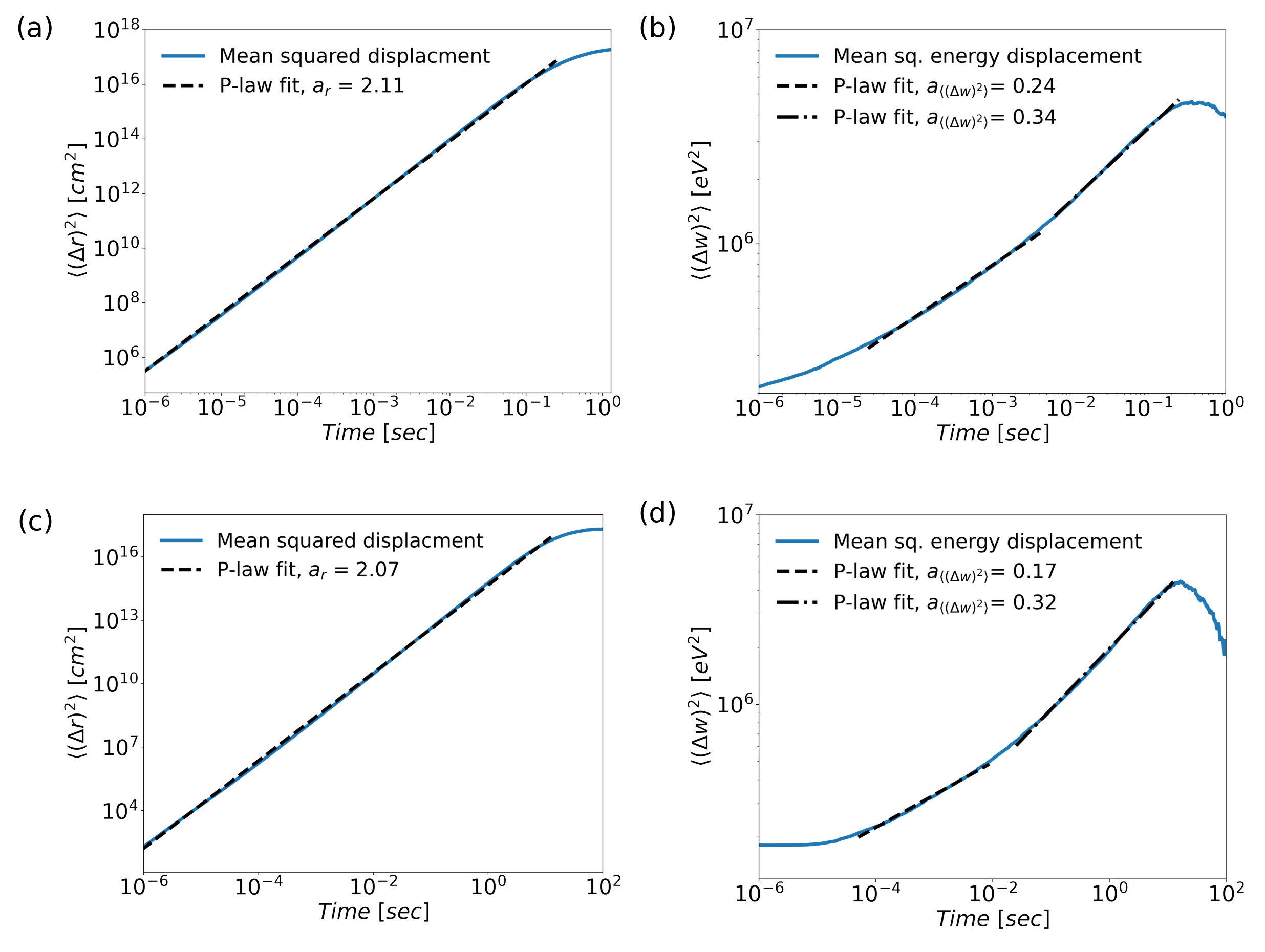}
         \caption {(a) Spatial mean squared displacement as a function of time for the electrons.  
         (b)  Mean squared displacement in energy as a function of time for the electrons. 
         (c) Spatial mean squared displacement as a function of time for the ions.  
         (d)  Mean squared displacement in energy as a function of time for the ions.
}\label{rmsd_non_rec_UCS}
     \end{center}
     \end{figure}

  Figs. \ \ref{rmsd_non_rec_UCS} a,b exhibit the mean squared displacement in space and in energy as a function of time for the electrons. They both attain a power-law scaling, the electrons are super-diffusive in position space with power-law index $a_{r} \ = \ 2.11$, and they are clearly sub-diffusive in energy space, with power-law index close to $0.25$.   
  Figs.\ \ref{rmsd_non_rec_UCS}c,d give the mean square displacements in space and energy for the ions, and the ions show the same scaling indices as the electrons.  What differs between the two populations is the time scale required for the ions to diffuse, which is, in general, an order of magnitude larger than the one required for the electrons. Taking a closer look at the initial phase of the energization process, one can see that the heavier ions need more time to gain energy and to efficiently begin diffusing in space. This observation is also in accordance with the fact that protons tend to stay longer inside the acceleration volume, resulting in a higher value for the escape time when compared to the electrons' one.
  
  \begin{figure}[!ht]
     \begin{center}

         \includegraphics[width=1\textwidth]{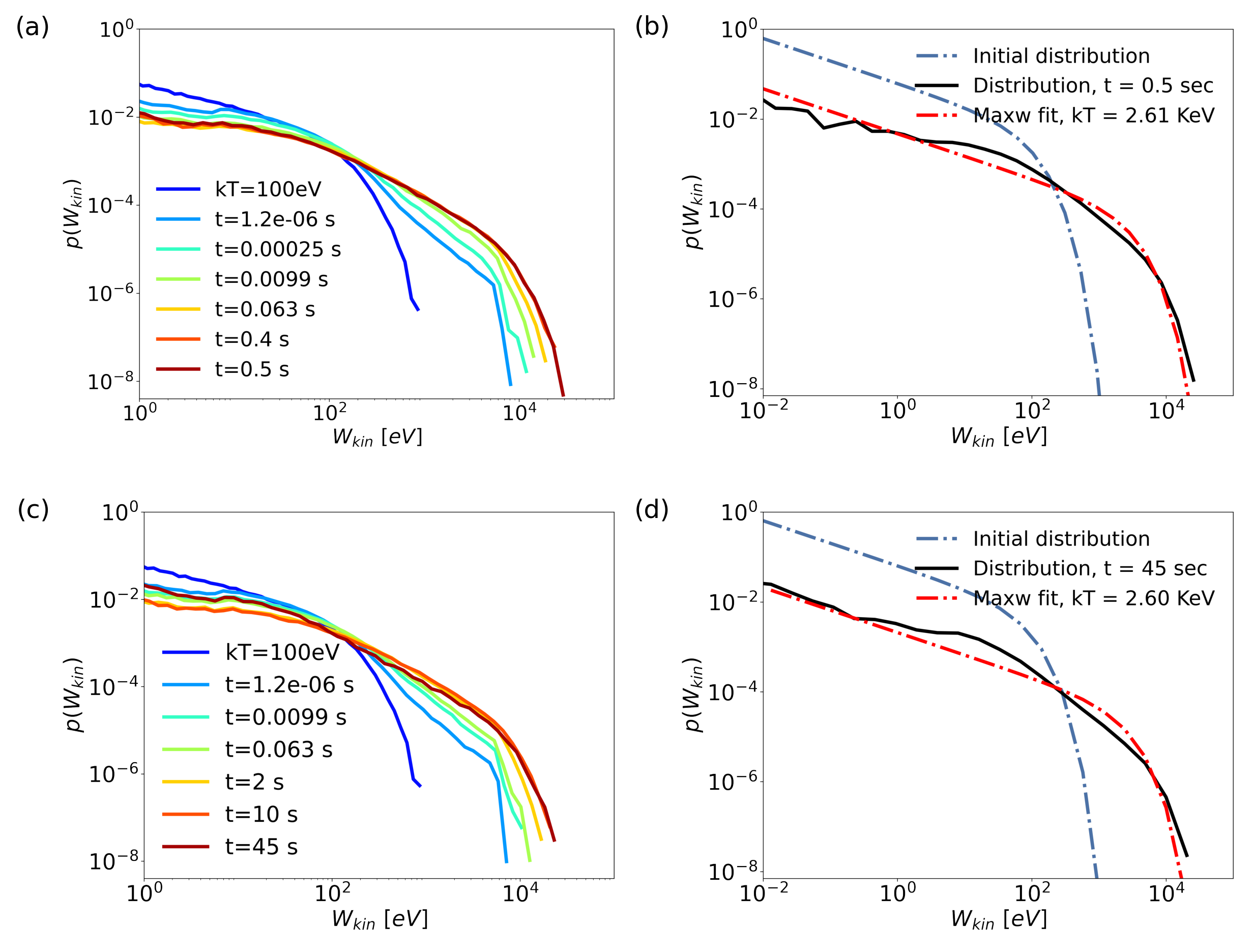}
       
       \caption { (a) The evolution of the kinetic energy distribution for the electrons.  (b) The electrons' steady state distribution at time $t\ = \ 0.5 \sec$, along with a Maxwellian fit giving a temperature $T \ = \ 2.61 \ keV$. (c) The evolution of the kinetic energy distribution for the protons. (d) The protons' steady state distribution at time $t\ = \ 45 \sec$, along with a Maxwellian fit giving a temperature $T \ = \ 2.60 \ keV$.
       }\label{non-rec-UCS_evol}
         \end{center}
     \end{figure}

  The delayed response of the protons to the external fields is also apparent when considering the evolution of the kinetic energy distribution, see Figs.\ \ref{non-rec-UCS_evol}a,c. Although in the steady-state the electrons' distribution does not have significant differences from the ions' distribution, the time scales needed to attain this very similar energization state differ significantly.

\begin{figure}[!ht]
     \begin{center} 
           
           \includegraphics[width=1\textwidth]{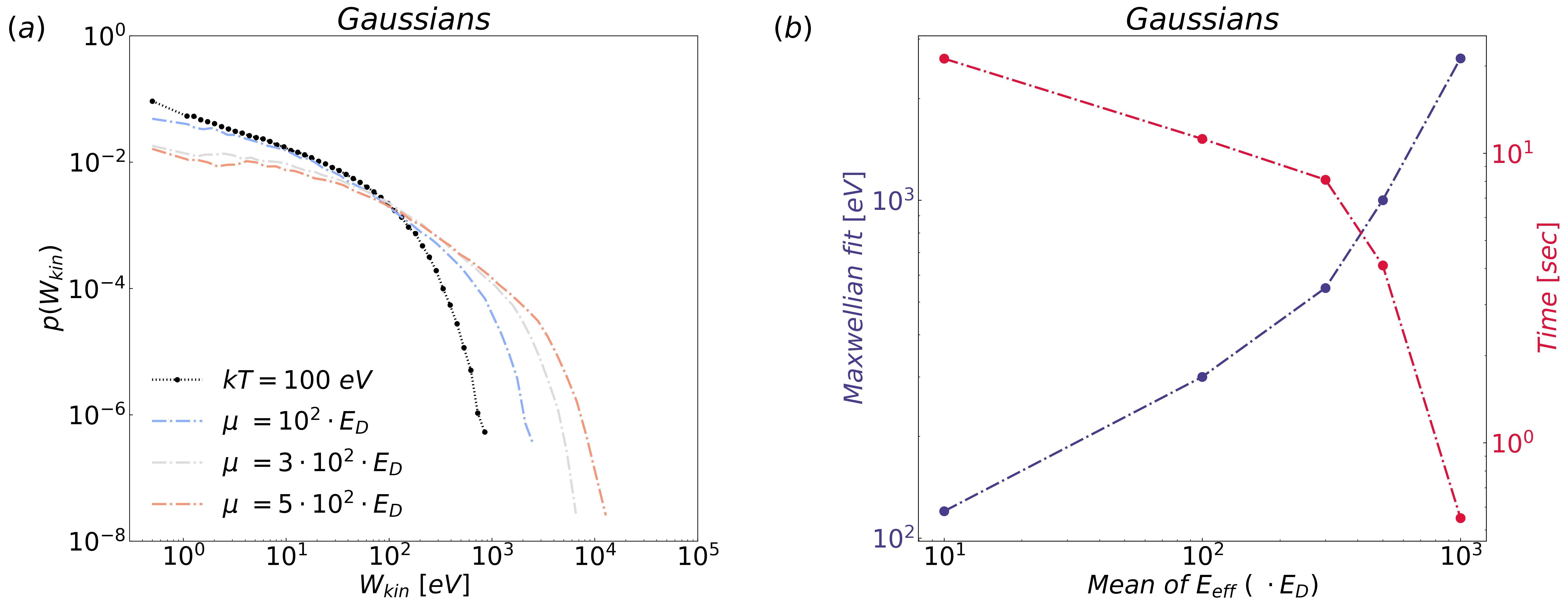}         
           \includegraphics[width=1\textwidth]{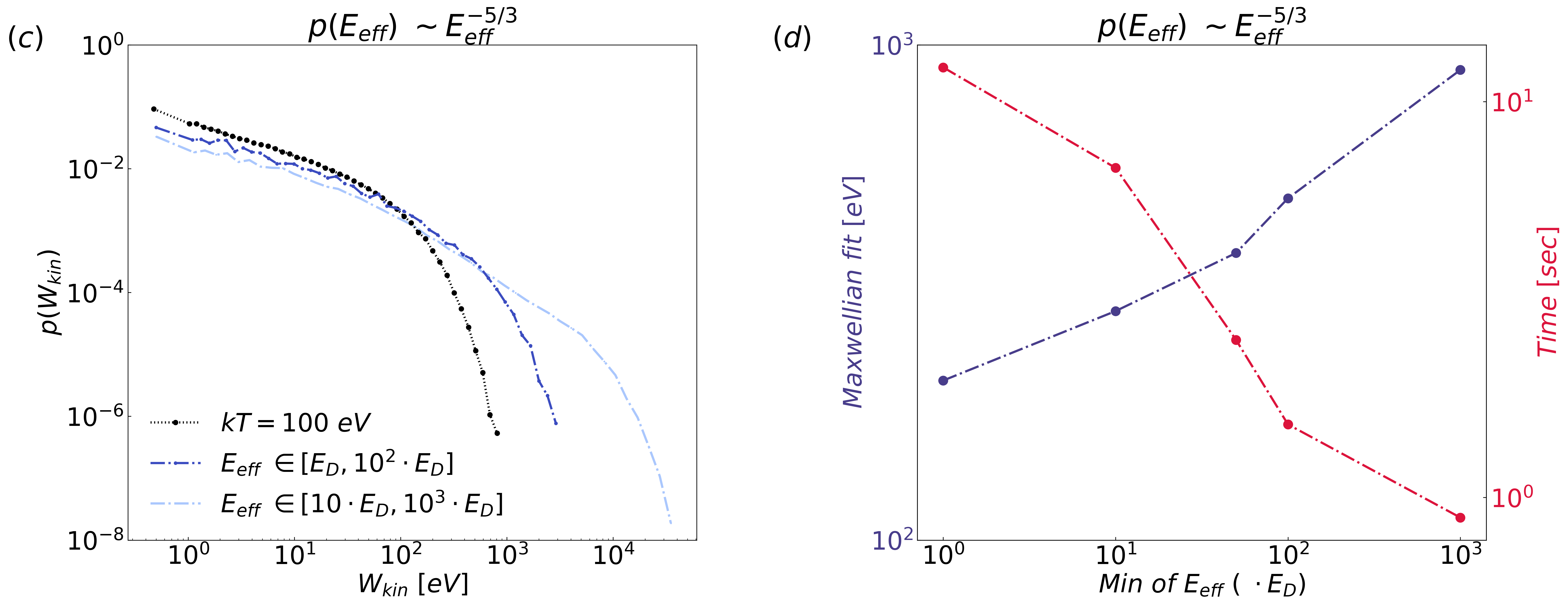}\\

         \caption {(a) The steady-state kinetic energy distribution for different Gaussian distributions of $E_{eff}$. (b) Parametric study of (i) the temperature of the Maxwellian fit (blue), and (ii) the time required for the distribution to reach steady state ($t_{acc}$; red) for Gaussian distributions of $E_{eff}$. (c) The steady-state kinetic energy distribution for different power-law distributions of $E_{eff}$. (d) Parametric study of (i) the temperature of the Maxwellian fit (blue), and (ii) the time required for the distribution to reach steady state ($t_{acc}$; red) for power-law distributions of $E_{eff}$. In any case, the maximum value of $E_{eff}$ is $E_{max} = 10^{2} \cdot E_{min}$. }\label{fig_final}
     \end{center} 
     \end{figure}

  Here, we have imposed a Gaussian distribution with mean ${\mu}_{E_{eff}} = \ 10^{3} \cdot E_{D}$ for the effective electric field $E_{eff}$, and the final kinetic energy distribution of both species in steady-state is close to a Maxwellian with temperature $\sim \ 2.6 \ KeV$ (see Figs.\  \ref{non-rec-UCS_evol}b,d), internally deformed to a power-law part, but no non-Maxwellian tail is developed. So the particles are only heated, and there is no acceleration. 
 The question of course is how the mean and standard deviation of the electric field's Gaussian distribution affect the steady-state kinetic energy distribution of the particles.  A related parametric study presented  in Fig. \ref{fig_final}a,b revealed the following:

\begin{itemize}
    \item {For any of the parameters tested, the steady-state distribution always is close to a Maxwellian (see Fig. \ref{fig_final}a). This means that imposing an effective electric field that follows a Gaussian distribution leads to an energization mechanism able to drive the particle distribution to higher temperatures. The mechanism, however, fails to account for particle acceleration, as no power-law tails are formed.}
    \item {As a result of increasing the mean of the Gaussian distribution to higher values, the kinetic energy distribution also attains higher temperatures, and the time needed to reach the steady-state is reduced (see Fig. \ref{fig_final}b).}
    \item {If the mean value of the Gaussian distribution is kept constant, but its standard deviation is reduced (i.e.\ the Gaussians become narrower), then the steady-state energy distribution fits better to a Maxwellian. In the limit where the Gaussian distribution is so narrow that it resembles a Delta function, the power-law part internal to the Maxwellian vanishes completely, and, depending on the mean of the imposed Gaussian, the kinetic energy distribution almost perfectly fits a Maxwellian .}

\end{itemize}

We have also performed a parametric study with a power-law distribution $P(E_{eff}) \sim E^{-5/3}$ for a number of lower ($E_{min}$) and upper ($E_{max}$) limits of the values of the electric field. In Fig.\  \ref{fig_final}c,d, we show the effects of this study on the electron steady-state energy distribution. We can conclude:
      \begin{itemize}
          \item{Keeping the limits within a relatively small order of magnitude apart, $\frac{E_{max}}{E_{min}} \ \sim \ 10^{2}$, the steady-state distribution closely resembles a Maxwellian for all of the cases tested (see the parametric study in Fig.\ \ref{fig_final}c).}
          \item{As a result of slightly adjusting the value of $E_{max}$ to higher values, while keeping  $E_{min}$ constant, the steady-state energy distribution starts to diverge from a Maxwellian distribution, as a short and steep power-law tail forms on top of the thermalized part (not shown in Fig.\ \ref{fig_final}c)}.
          Further increasing $E_{max}$ results in a progressively softer power-law tail that extends to higher values in the energy domain.  One has to keep in mind, however, the limitations imposed by the absence of reconnection on the physically attainable values of the electric field. It would be hard to imagine a too wide range in the values of electric fields in an environment dominated by non-reconnecting CS.

         \item{Adjusting $E_{min}$ to higher values drives the particle distribution to higher energies, reducing the time required to reach the steady-state (see Fig.\  \ref{fig_final}d).}
         
     \end{itemize}    
     
      Overall, we can conclude that for a power-law distribution of $E_{eff}$, the minimum value $E_{min}$ affects the thermalized part of the distribution, i.e.\ the steady state energy distribution becomes hotter for higher values of $E_{min}$. The shape of the distribution (i.e.\ how closely it resembles a Maxwellian) is determined by the ratio $\alpha_{range} \ =\frac{E_{max}}{E_{min}}$. For $\alpha_{range} \ \sim 10^{2}$ the steady state distribution is closely related to a Maxwellian. For larger values  of $\alpha_{range}$ the distribution starts to develop a power-law tail on top of the thermalized part, which gets increasingly softer as $\alpha_{range}$ increases.

\subsubsection{Non-reconnecting CS's as a possible coronal heating mechanism.}

 We have previously shown that stochastic energization of particles through non-reconnecting CSs proves to be a  mechanism that can effectively thermalize an injected particle distribution to increasingly high temperatures, depending on the mean value of the effective electric field imposed. Is this mechanism, however, able to drive a Maxwellian distribution of a typical chromospheric temperature of ${T \ = \ 10^{4} \ K}$ to the observed temperature of the solar corona of around ${T \ = \ 10^{6} \ K}$? To answer this question, we reduce the initial temperature of the particles to ${T \ = \ 0.85 \ eV}$. We also reduce the mean value of the imposed electric field distribution to ${\mu}_{Eff}\ = 10^{2} {\cdot} \ E_{D} $ and the standard deviation to ${\sigma}_{Eff}\ =  \ E_{D} $.

   \begin{figure}[!ht]
     \begin{center}
          \includegraphics[width=1\textwidth]{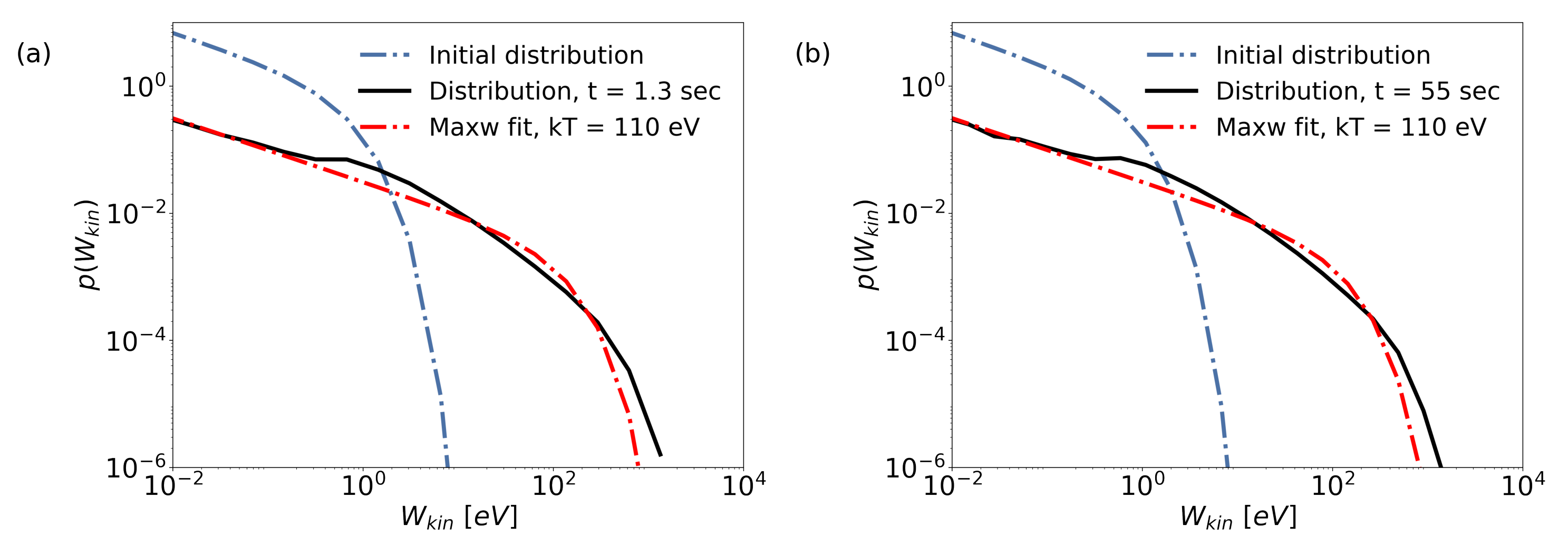} 
          
          \caption { The steady-state kinetic energy distribution for non{-}reconnecting CSs, where $E_{eff}$ follows a Gaussian distribution of mean ${\mu\ {=}\ 10^{2} {\cdot} E_{D}}$ and standard deviation  ${\sigma\ {=}\  E_{D}}$, (a) for electrons, (b) for protons, along with a Maxwellian fit of temperature $T \ = \ 110\ eV$ in both cases.
         }\label{coronal_heating}
     \end{center}
     \end{figure}

 Otherwise, we keep the same setup as described in Sec. \ref{non-rec-UCS}. In Fig.\ \ref{coronal_heating}a,b we show the steady-state kinetic energy distribution for the electron and proton populations, respectively. In both cases the steady-state distribution very closely resembles a Maxwellian distribution of temperature ${T \ = \ 110 \ eV \ {\sim} \ 10^{6} \ K}$. What differs between the two populations is the time needed for the distribution to reach its steady-state, which is of the order of a few seconds for the electrons, while it is of the order of tens of seconds for the protons. 
 
 \par
 Similar as in Sec.\ \ref{non-rec-UCS}, if we increase the mean value of the Gaussian distribution of the effective electric field $E_{eff}$, higher temperatures are achieved in the final distribution, while, at the same time, a slight decrease in the acceleration time is observed.

\subsection{Synergy of reconnecting and non{-}reconnecting  CSs.}\label{Synergy}

    Following the model described in Sec.\ \ref{sec:headings}, we assume that the scatterers inside the turbulent volume are divided into two classes, reconnecting and non-reconnecting CSs. A fraction $P$ ($0 \leq  P \leq 1$) are  CSs,  where the particles undergo stochastic acceleration, and the remaining fraction $1-P$ are RCSs that give stochastic energy kicks to the particles. The nature of a scatterer is determined by drawing a random number ${\alpha}$ in the range $[0,1]$, and if ${\alpha} \leq P$ we consider a non-reconnecting CS, otherwise a RCS.

\begin{figure}[!ht]
     \begin{center}
     \includegraphics[width=0.5\textwidth]{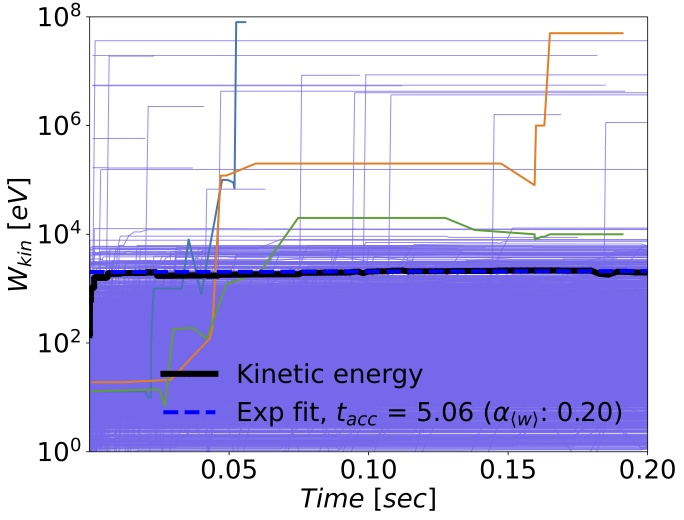}
        \caption {The mean energy as a function of time, and the energy of a few selected particles, marked by different colors. The particles undergo systematic and stochastic interactions with the CSs, with $P=0.5$.}
         \label{part_energ}
     \end{center}
     \end{figure}

 Regarding the characteristics of the non-reconnecting CSs, we follow the setup presented in Sec.\ \ref{non-rec-UCS}, and the statistical properties of the RCSs, as well as the characteristics of the acceleration volume, are kept as in Sec.\ \ref{UCS}. At time $t   =  0 $, a population of $10^{6}$ particles is injected into the mixed environment. The injected distribution is a Maxwellian of temperature $T  =  100 eV$. An extensive analysis has been performed for both electrons and ions, and the results presented here refer to electrons. The heating and acceleration characteristics of ions and electrons  interacting with the mixed environment of scatterers are similar, the only difference being the time scales required to reach a steady state.]

We first assume that $P\ = \ 0.5$, the probability to encounter a non-reconnecting CS is equal to the probability of an encounter with a RCS. The effect of combining a stochastic energization mechanism (non-reconnecting CSs) with a systematic one (RCSs) is presented in Fig.\ \ref{part_energ}, where there is a combination of a classical random walk like behavior with sudden and often rather large increases in energy, i.e.\ a kind of strongly positively biased random walk.


     \begin{figure}[!ht]
     \begin{center}
          \includegraphics[width=1\textwidth]{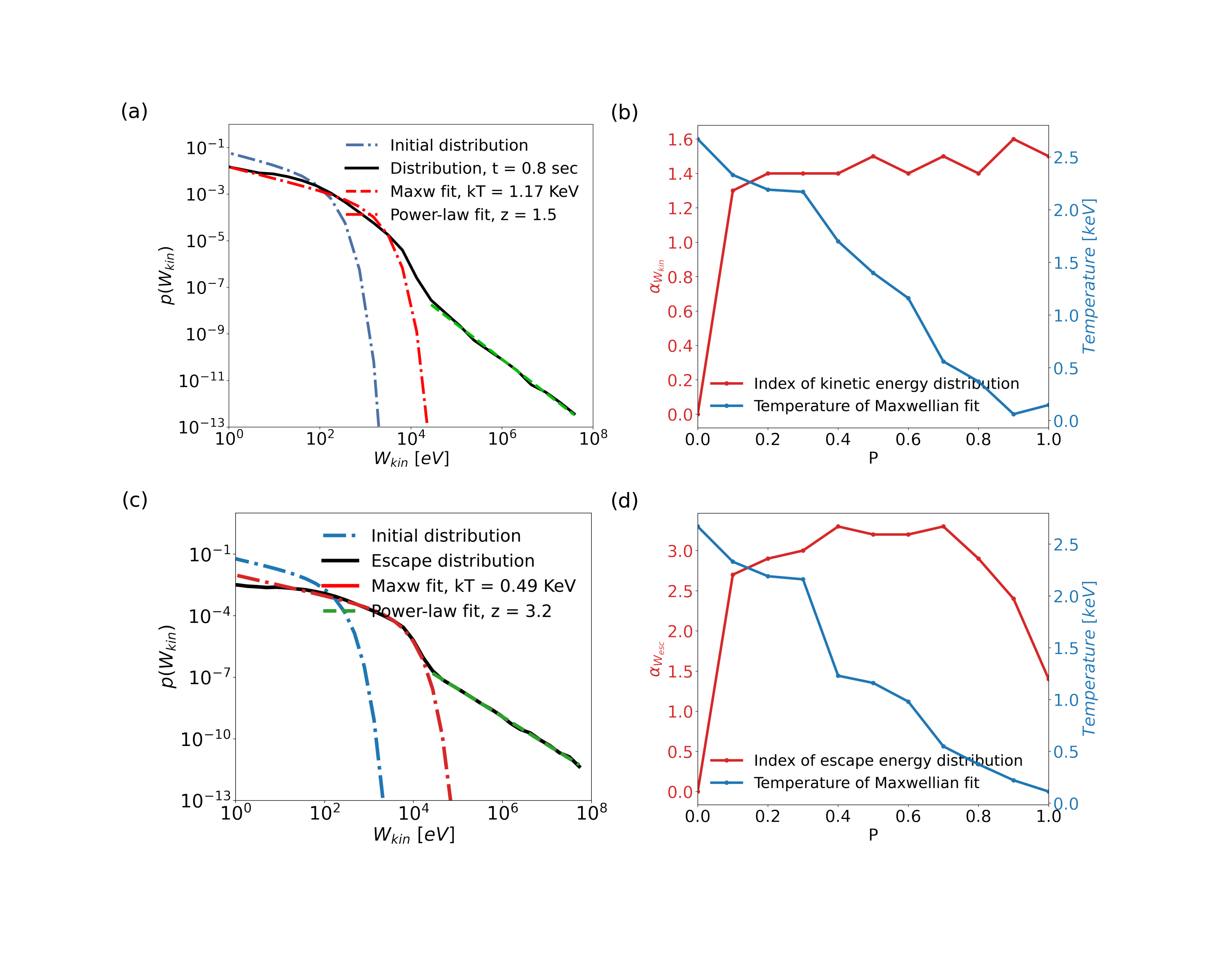} 
          
         \caption {(a) The steady-state kinetic energy distribution, along with a Maxwellian fit of temperature $T\sim 1\, KeV$ and a power-law fit, for $P {=} 0.5$.
         (b) Power-law index of the steady-state kinetic energy distribution
         (blue), and the temperature of the Maxwellian fit (red), for different ratios $P$ of the two kinds of scatterers. 
         (c) The kinetic energy distribution for the escaped particles, for $P=0.5$. 
         (d) Power-law index of the kinetic energy distribution of the escaped particles
         (blue), 
         and the temperature of the Maxwellian
         fit (red), for different ratios $P$ of the two kinds of scatterers.}\label{Wkinet_a}
     \end{center}
     \end{figure}

Since the fraction of scatterers acting as reconnecting or non-reconnecting current sheets in a turbulent environment cannot be expected to be always constant (e.g.\ it can vary as the system evolves in time), it is important to understand how different combination ratios of reconnecting to non-reconnecting CSs affect the energization process. Thereto, we start from the case $P =  0$, and we gradually increase the fraction of RCSs present in the system in our study.

 \begin{figure}[!ht]
     \begin{center}
          \includegraphics[width=0.5\textwidth]{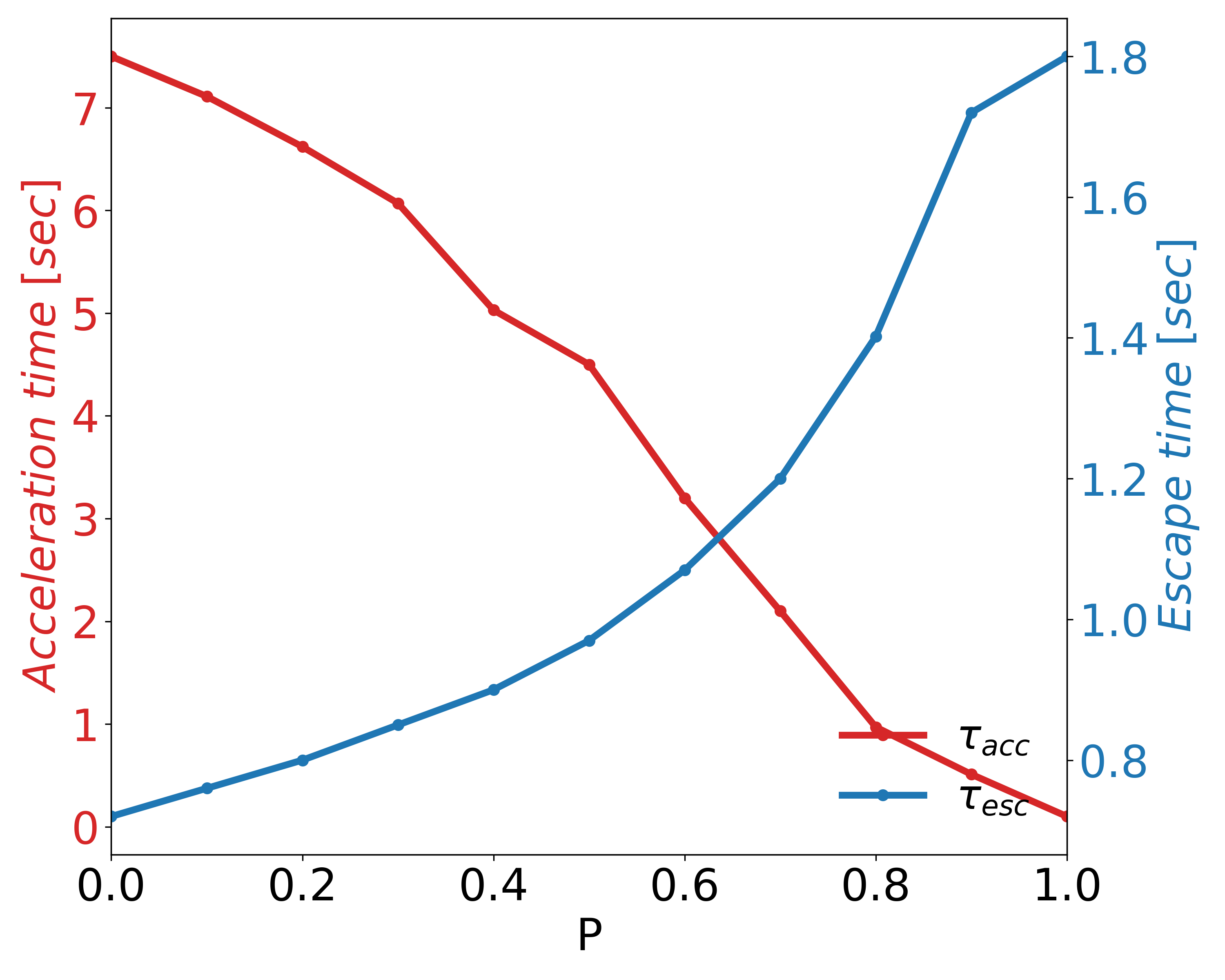}  
         
         \caption {Parametric study of (i) the escape time ($t_{esc}$), and (ii) the acceleration time ($t_{acc}$) for different ratios $P$ of the two kinds of scatterers.}\label{time}
     \end{center}
     \end{figure}
     
When examining the steady-state kinetic energy distribution of the particles, we observe a significantly different behavior at the two ends of the spectrum, going from distributions of minor heating and extended power-law tails for $P\ = \ 1$ (i.e.\ only reconnecting RCSs are present in the system) to almost perfect Maxwellians of higher temperature without a tail for $P =  0$, see Fig.\ \ref{Wkinet_a}b. In Fig.\ \ref{Wkinet_a}a, we present the steady-state kinetic energy distribution for the particles that remain inside the turbulent volume at time $t {\sim} 0.8 \, sec$, along with the injected distribution, a Maxwellian fit, and a power-law fit, for $P=0.5$. The energization process
heats the low-energy particles, where the distribution follows a Maxwellian with a temperature $T =  1 keV$. In the high energy part of the distribution, a power{-}law tail is formed, which, in steady-state, has  index $k=1.48$ and extends from about
$10\,$keV to $100\,$MeV. The power-law tail forms after a short period of a few milliseconds and persists even when more electrons have
escaped from the acceleration volume. We can therefore conclude that a combination of the two kinds of scatterers changes the behavior of the turbulent volume from either a pure systematic accelerator or a pure efficient heating mechanism, for the values $P =  1$ or $P=0$, respectively,  to a combined and efficient heating and acceleration mechanism. Very similar behavior can be observed when studying the energy distribution of the escaping particles, as shown in Figs.\ \ref{Wkinet_a}c,d.


Also important to note is the fact that as the percentage of RCSs present in the system increases, the acceleration time $t_{acc}$, which is proportional to the time needed for the kinetic energy distribution to attain its asymptotic state, decreases (see Fig.\ \ref{time}).  
The opposite behavior is observed when considering the escape time of the particles, i.e.\ $t_{esc}$ increases as the percentage of RCSs increases (see Fig.\ \ref{time}).

\begin{figure}[!ht]
     \begin{center}
     \includegraphics[width=0.5\textwidth]{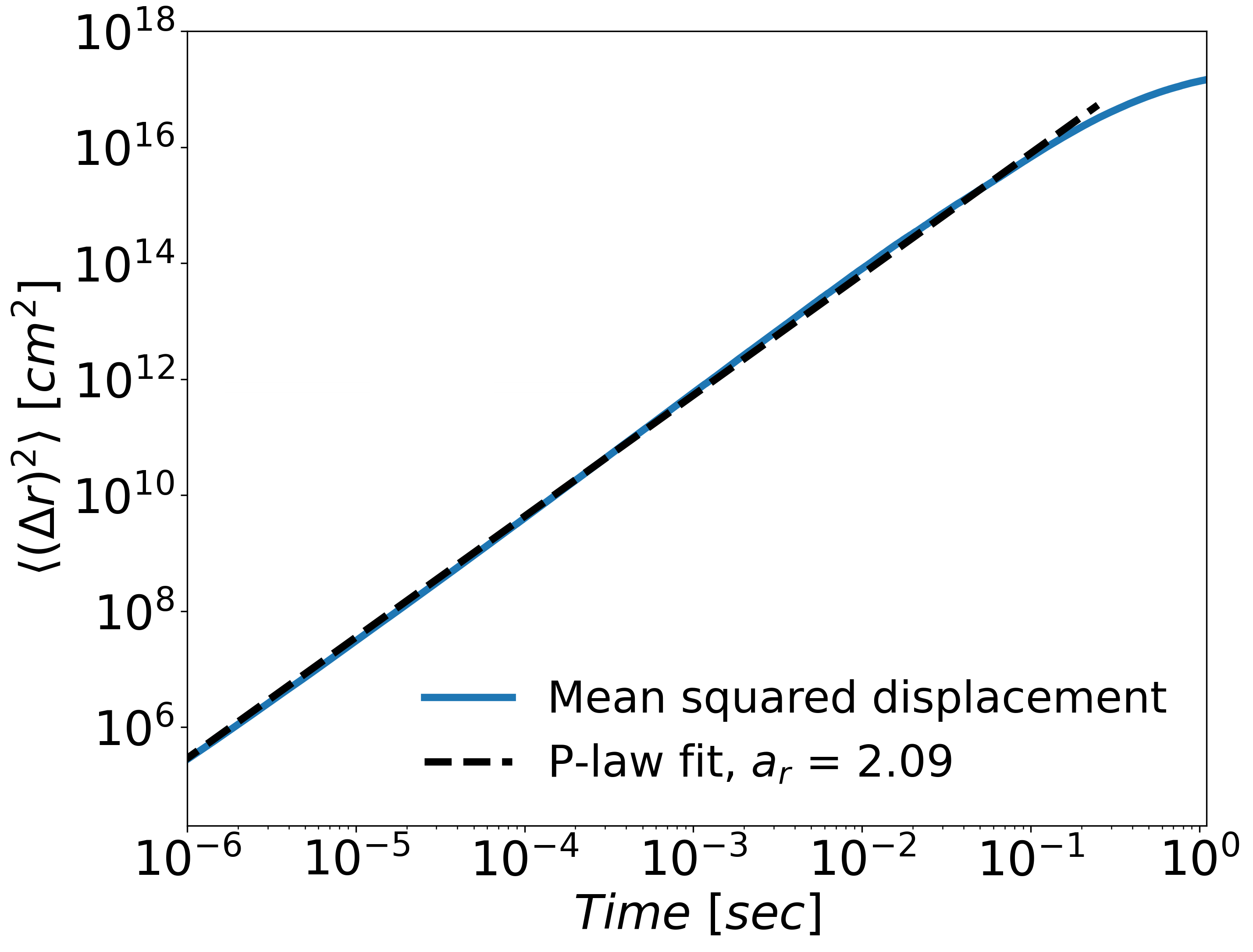}
         \caption {The  mean squared displacement as a function of time for $P =  0.5$.
         }\label{rmsd2}
     \end{center}
     \end{figure}
     


 In order to perform a study of the transport properties in space for the particle population, we once again make use of the set of predefined monitoring times described in Sec.\ \ref{sec:headings}.  In Fig.\ \ref{rmsd2}, we show the spatial mean squared displacement as a function of time, which follows a power-law scaling with index $k = 2.09$ for $P=0.5$. Also, the inset figure presents a parametric study for different ratios of reconnecting to non-reconnecting CSs. 
 Important to note is the fact that for $P  \geq 0.7$, the diffusion process has two stages, showing a ballistic behavior for times up to $t = 10^{-2}$ sec with $a_{r}$ ranging in $ [2.06 , 2.08]$, and a superdiffusive scaling for larger times with $a_r$  taking values in the range $ [1.53 , 1.68] $ (similarly to Fig.\ \ref{rmsd1}a). 
 For values $P  \leq 0.7$, $a_{r}$ practically remains constant, $a_{r} \ {\sim} \ 2.1$, throughout the diffusion process.


\section{Summary and conclusions}
 
 We have analyzed the energization of charged particles (electrons and protons) through their interaction with an environment of fractally distributed scatterers. The scatterers have been modeled to behave either as  RCSs, or non-reconnecting CSs. We have also tested a number of environments with the two energization mechanisms co-existing. 
 
 We have shown that the systematic interaction of particles with fractally distributed RCSs provides an efficient mechanism able to accelerate electrons and ions to very high energies, up to hundreds of MeV in a very short time. In the steady-state, the energy distributions exhibit very minor heating, and for energies, above $10^{3} \, eV$  there always is a power-law tail of index $z\ = 1.4-1.5$ for the particles staying inside the acceleration volume, whereas the distributions of the escaping particles show a power-law tail with index $\sim 1.4$. High energy particles undergo significant acceleration and easily depart from the energization volume. Their energy distribution is strongly correlated with the distribution of the energizations received from the RCSs. The extent of the power-law tail has been shown to be affected by the strength of the magnetic fluctuations  (${\delta}B$), and the maximum observed kinetic energy of the particles coincides with the maximum value of the energy increments ${\delta}W$, which are a function of ${\delta}B$. On the other hand, the kinetic energy distribution does not depend on the size of the simulation box.  
 An analysis of the spatial transport properties for electrons interacting with RCS shows that spatial diffusion is a two-stage process. A ballistic phase during the first 10$^{-2}\ sec$ is followed by a less impulsive superdiffusive phase lasting at least up to ${\sim}1 \sec$.
 Regarding the kinetic energy distribution, the results are identical for electrons and protons. What changes, however, is the increased time required for the protons to attain their steady-state distribution, as well as to spread through the simulation volume. 
 
 Replacing the scatterers with non-reconnecting CSs, we have shown that stochastic energization by non-reconnecting CS is a process that slowly but steadily heats particle distributions and increases their temperature, and it proves to be an efficient mechanism that can reproduce the heating of the Solar corona. The steady-state distribution's temperature strongly depends on the mean value of the imposed electric field (${\mu}_{E_{eff}}$). When ${\mu}_{E_{eff}}$ increases, the steady-state distribution attains higher temperatures, while at the same time the acceleration time slightly decreases. Proton and electron distributions show similar behavior. Protons, though, require much more time to reach steady-state distributions. Particle transport in any case is superdiffusive in position space and sub-diffusive in energy space.  Based on the analysis presented in this article, we suggest that coronal heating is based on non-reconnecting current sheets with a Gaussian distribution of the effective electric field of the CSs, which can be viewed, according to \cite{Einaudi21}, as another realization of the ``nano-flares" model proposed by \cite{Parker83}. We claim that "nanoflares" represent a small percentage of the CSs formed through the tangling of coronal magnetic field lines by the photospheric flows,  whose signature possibly are the "campfires" recently observed by the Solar Orbiter mission (\cite{Berghmans:5333, ChenY21}). We predict that when the density of "campfires" increases in the quiet corona, non-thermal particles will be present during coronal heating. Using a power-law distribution for the effective electric field $E_{eff}$, we reproduce the results obtained with the Gaussian distribution if  $a_{range}=E_{max}/E_{min}$ is relatively small. For large values of $a_{range}$, a very soft power-law tail appears in the energy distribution. Assuming that the non-reconnecting CSs are associated with relatively weak electric fields, we can claim that they dominate the heating of the solar plasma independently of the probability distribution function of their effective electric field.
 
 The effects of combining reconnecting with non-reconnecting CSs in the same environment have also been investigated. A parametric study on $P$ (i.e.\ the ratio of reconnecting to non-reconnecting CSs) has shown that steady-state energy distributions generally consist of power-law tails for the high energy particles on top of a thermal part at low energies. The degree of the distribution's heating, the extent and the index of the power-law tail, as well as the escape and the acceleration time, heavily depend on $P$. More specifically, when non-reconnecting CSs dominate, the distributions are heated to high temperatures while the extent of the power-law tail gets shorter i.e.\ the tail begins at higher energies). For environments with larger $P$, where  RCSs prevail, the steady-state distributions are characterized by minor heating and extended power-law tails of index $z\ = \ 1.4 - 1.5$.  The escape and acceleration time depend on the fraction of the two types of scatterers. The higher the fraction of RCSs is, the larger is the escape time, while at the same time the smaller is the acceleration time for the particles.

How turbulent reconnection sets in in the solar corona depends on the large-scale MHD instabilities driven by the unstable magnetic topologies, e.g.\ emerging magnetic flux, or kink instabilities. In most cases, a large-scale RCS sets in, which soon fragments, forming gradually a mixture of reconnecting and non reconnecting CSs. According to our analysis, the acceleration of high-energy electrons and ions dominates in the early phase when most CSs are still reconnecting, and then the gradual appearance of non-reconnecting CSs will heat the plasma inside the turbulent reconnecting volume. On the other hand, the random shuffling of the magnetic fields that emerged from the turbulent convection zone or the emergence of small-scale magnetic loops drives easily non-reconnecting CSs, which heat the solar corona and high energy particles are absent. 

Therefore, the solar corona operates in two main modes. Explosive events start with reconnecting CSs and gradually end up with a mixture of reconnecting and non reconnecting CSs (based on our analysis, this can be achieved with a ratio $P(t)$ varying in time). In the "quiet corona", the random shuffling of the magnetic fields dominates the evolution of the coronal magnetic field, whereby non-reconnecting CSs (elementary events \cite{Kanella17, Kanella18, Einaudi21}) heat the solar atmosphere and drive the confined flares \citep{Galsgaard96}. 

Our suggestions in this article depart from the currently very popular cartoon of the "standard flare scenario" and claim that each flare is a unique combination of non-reconnecting and reconnecting CSs hosted in the 3D complex magnetic topology that is driven by convection zone turbulence. We cannot unravel all this complexity by analyzing an isolated RCS and without taking into account the presence of coherent structures in a turbulent environment that hosts RCSs and non-reconnecting CSs. 

The innovative parts of our work are (1) the use of a realistic 3D turbulent reconnection volume, with a fractal distribution of CSs, (2) the estimate of the energization and escape times inside the turbulent volume, (3) the investigation of the role of the systematic, the stochastic and the synergy of both energization mechanisms inside the turbulent reconnection volume, (4) the estimate of the energization (heating and acceleration) of electrons and ions, and their transport properties in energy and space. The weak points of our analysis are related to the fact that we use test particle simulations in steady-state turbulent reconnection environments, so the electromagnetic fields do not evolve in time and there is no feedback from the particles to the fields.  At least a time-dependent evolution of the ratio of the RCSs ($P(t)$) would be needed to realistically explore the evolution of explosive events. Also, the statistical characteristics of the reconnecting and non-reconnecting CSs are not known. We hope that the results from the Parker Solar Probe and future 3D MHD simulations, which follow the formation and evolution of turbulent reconnection volumes in time and space and incorporate the feedback from the kinetic evolution of the energetic particles in explosive and non-explosive events, will clarify these issues soon.

\begin{acknowledgements}
We thank the anonymous referee for his comments and suggestions, which helped us to improve the article. Nikos Sioulas was supported by the  HERMES DRIVE NASA Science Center grant No. 80NSSC20K0604. 
     \end{acknowledgements}

\end{document}